\documentclass[12pt,twoside]{article}

\date{}
\def\aap{A\&A\,  }
\def \aapr{The Astronomy and Astrophysics Review}
\def\aj{AJ  }
\def\apj{ApJ\,  }
\def\apjl{ApJ\,  }
\def\apss{Astrophysics and Space Science  }
\def\cjaa{Chinese J. Astron. Astrophys.  }

\def\mnras{MNRAS\,  }

\def\pasj{PASJ\,  }

\usepackage{url}
\usepackage {graphicx}
\usepackage {amsmath}
\def\ca{\rm{CA}\,{\small\rmfamily II}\,}
\def\snr{SN \,1993J\,}
\def\sn1987a{SN \,1987A\,}
\def\s1006{SN \,1006\,}
\def\firstnova{SN \,1604\,}
\def\duesnr{SN\, 2009ig~}
\def\velu{\rm\,{kms^{-1}}}
\def\sun{\hbox{$\odot$}}
\def\degreezan{^{\,\circ}}
\DeclareRobustCommand{\orderof}{\ensuremath{\mathcal{O}}}
\begin{document}

\centerline{\bf ADVANCES IN APPLIED PHYSICS, Vol. x, 20xx, no. xx, xxx - xxx}

\centerline{\bf HIKARI Ltd, \ www.m-hikari.com}

\centerline{\bf http://dx.doi.org/10.12988/}

\centerline{}

\centerline{}

\centerline {\Large{\bf
The physics of asymmetric supernovae and 
}}

\centerline{}

\centerline{\Large{\bf
supernovae remnants
  }}

\centerline{}

\centerline{\bf {L. Zaninetti}}

\centerline{}

\centerline{Physics Department,}

\centerline{Universit\`a degli Studi di Torino,}

\centerline{via P. Giuria 1,  10125 Torino, Italy}

\centerline{}
{\footnotesize Copyright
$\copyright$ 2015 Zaninetti Lorenzo.
This is an open access article distributed
under the Creative Commons Attribution License,
which permits unrestricted use,
distribution, and reproduction in any medium,
provided the original work is properly cited.}
\begin {abstract}
We model the circumstellar medium 
with four density profiles:
hyperbolic  type,
power law   type,
exponential type
and Gaussian type.
We solve analytically or numerically 
the four first-order differential equations
which arise in the framework of the classical 
thin layer approximation.
The  non-cubic dependence of the swept mass 
with the advancing radius is also considered.
We derive the equation of motion for 
the thin layer approximation in special relativity in two cases.
The initial conditions are chosen in order to model
the temporal evolution of SN 1987A   over 23  years
and of SN 1006 over 1000 years.
We review the building blocks of the
symmetrical and asymmetrical formations of the image.
\\
PACS keywords                    \\
{
supernovae: general
supernovae: individual (SN 1987A)
supernovae: individual (SN 1006)
ISM       : supernova remnants
}
\end{abstract}

\section{Introduction}

The term \textit{nova} derives from the Latin \textit{novum},
which means new.
As an example, a `Stella Nova', now \firstnova,   
was observed  
both by Kepler in Italy, see \cite{Kepler1606},
and Galileo, see
\cite{Favaro1890,Shea2005}
as well as by Korean astronomers, 
see \cite{Clark1977,Ruiz-Lapuente2017}.
 
More recent is the use of the term  \textit{supernova} (SN), 
see \cite{Gaposchkin1936}. 
The  expanding shell of matter from an SN,
consisting of the supernova ejecta
and ‘swept-up’  matter, 
has been called an SN remnant (SNR), see \cite{Brown1954}.
Astronomers initially 
thought  that the equation for an SN was symmetric,
and  the following two laws of motion, among others, 
were used.
In the  Sedov solution, the radius, $r$,
scales  as  $r \propto t^{0.4}$,
where $t$ is the time, see \cite{Sedov1959,Dalgarno1987}.
In the  momentum conservation  for   a thin layer
approximation the radius 
scales  as  $r \propto t^{0.25}$,
see \cite{Dyson1997,Padmanabhan_II_2001}. 
Later the following two terms 
appeared: non-spherical SNR,   see \cite {Bodenheimer1984},
and symmetry in SN explosions, see
\cite{Amnuel1976}. 
The observed asymmetries can be classified 
when the external surface is rotated in front of the observer.
Often an equatorial or reflectional  symmetry  
is visible in the astronomical images. 
This is the case with  
the bipolar planetary nebula He 2-104, see \cite{Smith2005},
the protoplanetary nebula M2-9, see \cite{Scarrott1993},
the young planetary nebula MyCn 18, see \cite{Bryce2004}
and  
the nebula around Eta Carinae, see \cite{Smith2000}.
As an example  of  a rotation which brings the asymmetrical
object face on,  see 
Figure 1 in \cite{Smith2006}.
The magnetic fields around planetary nebulae 
have  been observed by \cite{Khesali2006,Sabin2007} 
and this observational fact has triggered
simulations of the asymmetrical shape of planetary
nebulae  based on variations of the magnetic field, 
see \cite{Khesali2006,Greaves2002,Sabin2007}.
In the following, the terms `asymmetrical', `asymmetric', `non-symmetric',
and `aspherical', are supposed to have  an identical meaning, i.e. 
they are 
synonyms.
The SNRs can therefore  be classified in
 light  of the observed symmetry.
A first example is  \snr,
which
presented
a
circular symmetry
for 4000
days, see \cite{Marcaide2009}.
An example of a weak departure from circular symmetry
is \s1006,  in which a ratio of 1.2
between the maximum and minimum radius
has been measured, see \cite {Reynolds1986}.
An example  of axial symmetry  is \sn1987a,
in which three rings
are symmetric  with respect to a line
which  connects the centres,
see  \cite{Tziamtzis2011}.
We now briefly review some results on the expansion 
velocity of SNs.
The  spectropolarimetry
(\ca IR triplet) of SN 2001el
gives a maximum velocity of $\approx 26000\velu$,
see  \cite{Wang2003}.
The same  triplet when searched in seven SNs
of type Ia gives
$10400 \velu  \leq v \leq 17700 \velu $,
see Table I in \cite{Mazzali2005}.
A time series of eight  spectra in \duesnr
allows asserting that the velocity at the  \ca line,
for example,
decreases  in  12 days from 32000 $\velu$ to 21500$\velu$,
see Fig. 9 in \cite{Marion2013}.

A recent analysis of 58 type Ia SNs in Si II
gives  $9660 \velu  \leq v \leq 14820 \velu $,
see Table II
in \cite{Childress2014}.
Other examples for the maximum velocity of expansion 
are 
$\approx 3700\, \velu$ , 
see Fig.~20.21 in \cite{Branch2017} and Fig.~6 
in \cite{Silverman2015}.
The  previous analysis  allows  saying that
the maximum velocity so far observed for an SN
is  $\frac{v}{c} \approx  0.123$, where
$c$ is the speed of light; this  observational fact
points to a relativistic equation of motion.

The above arguments  leave some questions
unanswered or only partially answered:
\begin {itemize}
\item 
Is it possible to deduce an  equation of motion
in the framework of the thin layer approximation 
adopting different density profiles?
\item 
Is it possible to deduce an equation of motion
for an expanding shell  assuming that
only a fraction of the mass
enclosed in the advancing sphere
is absorbed in the thin layer?
\item 
Is it possible to model the complex
three-dimensional (3D)
behaviour of
the velocity field of an expanding nebula?
\item   Is it possible to make an
       evaluation of the reliability of the
       numerical results
       for the radius, compared to the
       observed values?
\item   Can  we reproduce  complicated features,
       such as the equatorial ring + two outer
       rings  in \sn1987a, which are classified
       as `Mysterious Rings', see
       \cite{mystery}?

\item   Can  we reproduce  the appearance  of the rim 
       of \s1006, sometimes classified as
       `A contrail from an alien spaceship' or  
       `A jet from a black hole',
       see \cite{alien}.

\item   Is it  possible to
       build
       cuts of the model intensity which can
       be compared
       with existing
       observations?
\end{itemize}

\section{The density profiles}

\label{secprofiles}

This  section outlines   
the meaning of section for  a 3D surface,
the adopted symmetries, 
and then  postulates the existence of  four   density profiles
for the circumstellar medium (CSM) around the SNs:
an hyperbolic  profile,
a  power law   profile,
an exponential profile,
and 
a  Gaussian    profile.
The above profiles produce a reflectional symmetry 
about the Cartesian plane $z=0$.
We briefly recall  that  the density of  the galactic H\,I  
scales  with a Gaussian dependence  as 
\begin{equation}
n_H(z;b) = n_H(0) \exp({-\frac{z^2}{2 b^2}})
\quad ,
\label{galactich}
\end{equation}
where $z$ is the vertical distance in pc from the galactic plane,
$ n_H(0)=1.11$ $\mathrm{particles~}{\mathrm{cm}^{-3}}$,
$b=75.5\,pc$, and $z<1000\,pc$  see \cite{Zhu2017}.
In the  following we will explore gradients in density
with  smaller values in respect to  the scale $h$ of 
the galactic H\,I and we will see which gradient 
produces the best results.  

\subsection{Geometrical section of a 3D surface}

\begin{figure*}
\begin{center}
\includegraphics[width=7cm]{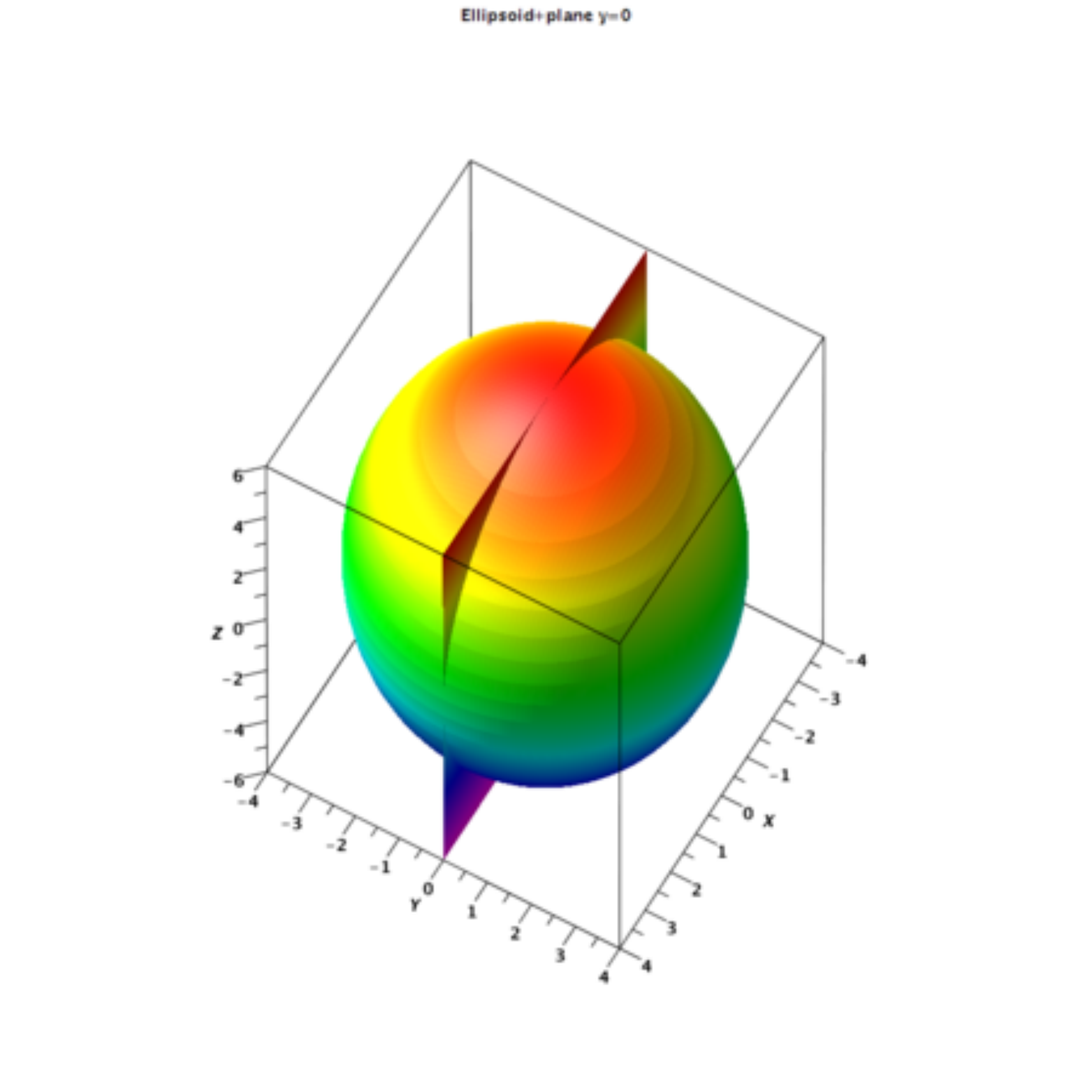}
\end {center}
\caption
{
An ellipsoid  
when $a=4 ,b=4$,  and $c=6$,   and the plane $y=0$.
}
\label{ellipsoid_plane}
    \end{figure*}

\begin{figure*}
\begin{center}
\includegraphics[width=7cm]{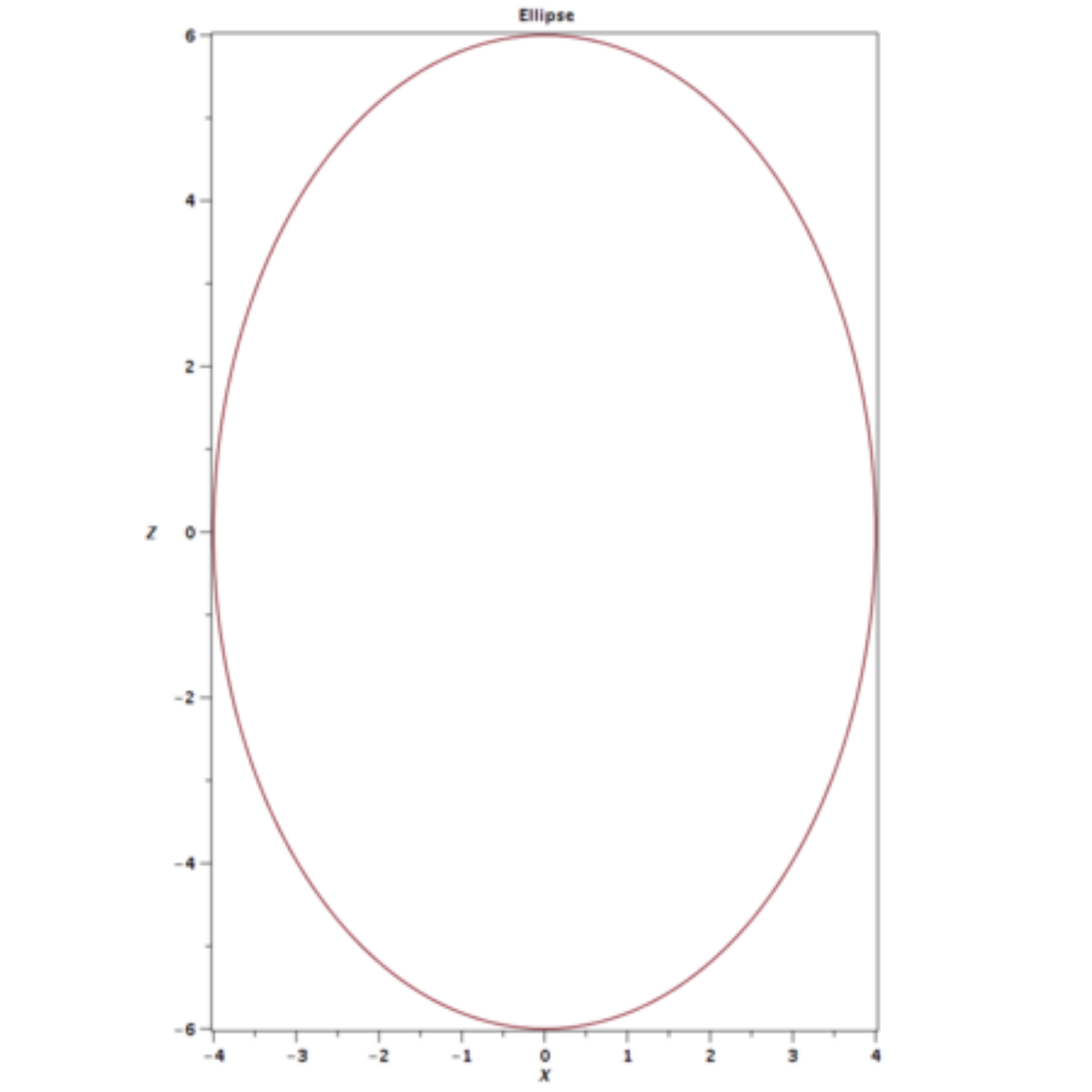}
\end {center}
\caption
{
The ellipse originating from the intersection of an ellipsoid with a plane.}
\label{ellipse}
    \end{figure*}

As an example, we take an  ellipsoid in Cartesian coordinates 
which has the equation 
\begin{equation} 
\frac{z^2}{a^2} + \frac{x^2}{b^2} + \frac{y^2}{c^2}=1 
\quad ,
\label{ellipsoid}
\end{equation}
where the lengths $a,b$, and $c$ are the semi-axes.
The  intersection  with the plane $y=0$ 
is displayed in 
Figure \ref{ellipsoid_plane}
and Figure \ref{ellipse} displays the resulting section,
which is an ellipse.

In the following, the meaning of geometrical section will be associated
with the intersection between a complex 3D surface
and the plane  $x=0$ or $y=0$.

\subsection{Spherical coordinates}

\begin{figure*}
\begin{center}
\includegraphics[width=7cm]{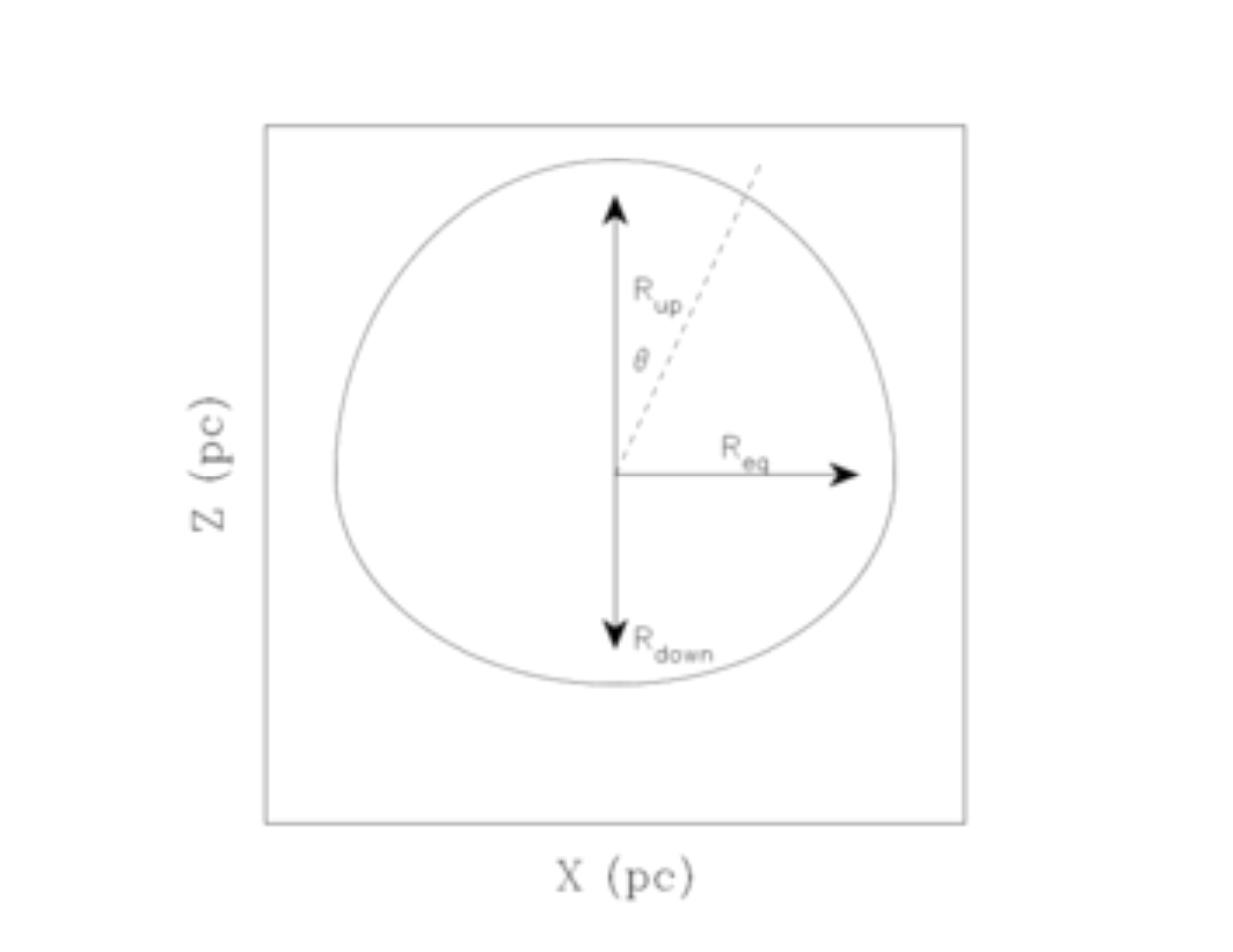}
\end {center}
\caption
{
Geometrical section for an asymmetric SN.
}
\label{section_asymm}
    \end{figure*}

A point in  Cartesian coordinates is characterized by
$x,y$, and $z$  
and  the position of the origin 
is the center of the SN explosion.
The same point in spherical coordinates is characterized by
the radial distance $r \in[0,\infty]$,
the polar angle     $\theta \in [0,\pi]$,
and the azimuthal angle $\varphi \in [0,2\pi]$.
Figure \ref{section_asymm} presents a geometrical section  of an
asymmetric SN in which is  clearly seen
the polar  angle
$\theta$ and the three
observable radii $R_{up}$, $R_{down}$, and $R_{eq}$.
We now outline two methods to measure the asymmetries.
The {\it first} method  evaluates  
the ratio $\frac{R_{up}}{R_{eq}} $
which models  the difference  between the observed radius in the polar 
direction in respect to the equatorial direction.
We can therefore speak of weak observed asymmetry
when  $1 < \frac{R_{up}}{R_{eq}} <2 $
and  great observed asymmetry
when  $\frac{R_{up}}{R_{eq}} >2  $.
As an example, the efficiency along the equatorial direction
is  
\begin{equation}
\epsilon_{\mathrm {eq}}  =100 \times \frac{|R_{eq}- R_{eq,theo}|}{R_{eq}}
\quad one\,direction 
\quad,
\label{efficiencyone}
\end{equation}
where $R_{eq,theo}$ is the theoretical radius given by the model along the 
equatorial direction.
 
The {\it second}  method  
is connected with the 
availability of  observed and theoretical 
geometrical sections of SN/SNR.
An observational
percentage reliability, $\epsilon_{\mathrm {obs}}$,
is  introduced over the whole range
of the polar   angle  $\theta$,
\begin{equation}
\epsilon_{\mathrm {obs}}  =
100(1-\frac{\sum_j |r_{\mathrm {obs}}-r_{\mathrm{num}}|_j}{\sum_j
{r_{\mathrm {obs}}}_{,j}})
\quad many\,directions 
\quad,
\label{efficiencymany}
\end{equation}
where
$r_{\mathrm{num}}$ is the theoretical radius given by a model,
$r_{\mathrm{obs}}$ is the observed    radius, and
the  index $j$  varies  from 1 to the number of
available observations.
The above statistical method allows fixing  the parameters
of the theory in a scientific way rather
than adopting an `ad hoc' hypothesis.

\subsection{A hyperbolic  profile}

The density  is  assumed to have the following
dependence on $z$
in Cartesian coordinates,
\begin{equation}
 \rho(z;z_0,\rho_0) =
  \begin{cases}
    \rho_0                 & \quad \text{if } z   \leq z_0\\
    \rho_0 \frac{z_0}{z}   & \quad \text{if } z   >    z_0\\
  \end{cases}
  \label{profhyperbolic}
\quad ,
\end{equation}
where the parameter $z_0$ fixes the scale and  $\rho_0$ is the
density at $z=z_0$.
In spherical coordinates
the dependence  on the polar angle is
\begin{equation}
 \rho(r;\theta,z_0,\rho_0) =
  \begin{cases}
    \rho_0                 & \quad \text{if } r \cos(\theta)  \leq z_0\\
    \rho_0 \frac{z_0}{r \cos(\theta)}   & \quad \text{if } r \cos(\theta)  >    z_0\\
  \end{cases}
  \label{profhyperbolicr}
  \quad  .
\end{equation}

Given a solid angle  $\Delta \Omega$
the mass $M_0$ swept
in the interval $[0,r_0]$
is
\begin{equation}
M_0 =
\frac{1}{3}\,\rho_{{0}}\,{r_{{0}}}^{3} \Delta \Omega
\quad .
\end{equation}
The total mass swept, $M(r;r_0,z_0,\alpha,\theta,\rho_0) $,
in the interval $[0,r]$
is
\begin{equation}
M(r;r_0,z_0,\alpha,\theta,\rho_0)= \bigl (
\frac{1}{3}\,\rho_{{0}}{r_{{0}}}^{3}+\frac{1}{2}\,{\frac {\rho_{{0}}z_{{0}} \left( {r}
^{2}-{r_{{0}}}^{2} \right) }{\cos \left( \theta \right) }}
\bigr) 
\Delta \Omega
\quad .
\label{masshyperbolic}
\end{equation}
The density $\rho_0$ can be  obtained
by introducing  the number density, $n_0$, expressed  in particles
$\mathrm{cm}^{-3}$,
the mass of  hydrogen, $m_H$,
and  a multiplicative factor $f$,
which is chosen to be  1.4, see \cite{Dalgarno1987},
\begin{equation}
\rho_0  = f  m_H n_0
\quad .
\end{equation}
The astrophysical version of the total swept mass,
expressed in solar mass
units, $M_{\sun}$,  is therefore
\begin{equation}
M (r_{pc};z_{0,pc},n_0,\theta)\approx
\frac{0.0172 \,n_{{0}}{   z}_{{{   0,pc}}}{r_{{{   pc}}}}^{2}}
{cos(\theta)}
\Delta \Omega \,M_{\sun}
\quad ,
\end{equation}
where
$z_{0,pc}$, $r_{0,pc}$ and $r_{0,pc}$
are  $z_0$,      $r_0$      and $r$ expressed  in pc.

\subsection{A power law profile}

The density  is  assumed to have the following dependence on $z$
in Cartesian coordinates:
\begin{equation}
 \rho(z;z_0,\rho_0) =
  \begin{cases}
    \rho_0                            & \quad \text{if } z   \leq z_0\\
    \rho_0 (\frac{z_0}{z})^{\alpha}   & \quad \text{if } z   >    z_0\\
  \end{cases}
  \label{profpower}
\quad ,
\end{equation}
where $z_0$ fixes the scale.
In spherical coordinates,
the dependence  on the polar angle is
\begin{equation}
 \rho(r,\theta,z_0,\rho_0) =
  \begin{cases}
    \rho_0                                         & \quad \text{if } r \cos(\theta)  \leq z_0\\
    \rho_0 (\frac{z_0}{r \cos(\theta)})^{\alpha}   & \quad \text{if } r \cos(\theta)  >    z_0\\
  \end{cases}
  \label{profpowerz}
\end{equation}
The mass $M_0$ swept
in the interval  $[0,r_0]$ in a given solid angle
is
\begin{equation}
M_0 =
\frac{1}{3}\,\rho_{{0}} \,{r_{{0}}}^{3} \Delta \Omega
\quad .
\end{equation}
The total mass swept,
$M(r;r_0,\alpha,\theta,\rho_0) $,
in the interval $[0,r]$
is
\begin{multline}
M (r;r_0,\alpha,z_0,\theta,\rho_0)=  \\
\Bigg (
\frac{1}{3}\,\rho_{{0}}{r_{{0}}}^{3}-{\frac {{r}^{3}\rho_{{0}}}{\alpha-3}
 \left( {\frac {z_{{0}}}{r\cos \left( \theta \right) }} \right) ^{
\alpha}}+{\frac {\rho_{{0}}{r_{{0}}}^{3}}{\alpha-3} \left( {\frac {z_{
{0}}}{r_{{0}}\cos \left( \theta \right) }} \right) ^{\alpha}}
\Bigg )
 \Delta \Omega
\quad .
\label{sweptmasspower}
\end{multline}
The astrophysical  swept mass is
\begin{equation}
M (r_{pc};z_{0,pc},\alpha,n_0,\theta )
\approx
\frac
{
0.03444\,n_{{0}}{{\it z}_{{{\it 0,pc}}}}^{\alpha}{r_{{{\it pc
}}}}^{-\alpha+3} \left( \cos \left( \theta \right)  \right) ^{-\alpha}
}
{
3 -\alpha
}
\Delta \Omega \,M_{\sun}
\quad ,
\end{equation}
where
     $z_{0,pc}$  and $r_{pc}$
are  $z_0$       and $r$ expressed  in pc.

\subsection{An exponential profile}

The density  is  assumed to have the following
exponential dependence on $z$
in Cartesian coordinates:
\begin{equation}
 \rho(z;b,\rho_0) =
\rho_0  \exp{(-z/b)}
\quad ,
\label{profexponential}
\end{equation}
where $b$ represents the scale.
In spherical coordinates,
the density is
\begin{equation}
 \rho(r;r_0,b,\rho_0) =
  \begin{cases}
    \rho_0                                 & \quad \text{if } r    \leq r_0\\
    \rho_0 \exp{-\frac{r\cos(\theta)}{b}}  & \quad \text{if } r    >   r_0\\
  \end{cases}
  \label{profexponentialr}
\end{equation}
The total mass swept,   $M(r;r_0,b,\theta,\rho_0) $,
in the interval $[0,r]$ is
\begin{multline}
M (r;r_0,b,\theta,\rho_0)= \\
\Bigg (
\frac{1}{3}\,\rho_{{0}}{r_{{0}}}^{3}-{\frac {b \left( {r}^{2} \left( \cos
 \left( \theta \right)  \right) ^{2}+2\,rb\cos \left( \theta \right) +
2\,{b}^{2} \right) \rho_{{0}}}{ \left( \cos \left( \theta \right) 
 \right) ^{3}}{{\rm e}^{-{\frac {r\cos \left( \theta \right) }{b}}}}}
\\
+
{\frac {b \left( {r_{{0}}}^{2} \left( \cos \left( \theta \right) 
 \right) ^{2}
+2\,r_{{0}}b\cos \left( \theta \right) +2\,{b}^{2}
 \right) \rho_{{0}}}{ \left( \cos \left( \theta \right)  \right) ^{3}}
{{\rm e}^{-{\frac {r_{{0}}\cos \left( \theta \right) }{b}}}}}
\Bigg )
\Delta \Omega
\quad .
\label{massexponential}
\end{multline}
The astrophysical version expressed in solar masses is
\begin{multline}
M (r;r_{0,pc},b_{pc},\theta,n_0)=  \\
\Bigg (
\frac{1}{ \left( \cos \left( \theta \right)  \right) ^{3}}
\times
\bigg (  0.01148\,{{\it r}_{{{\it 0,pc}}}}^{3} \left( \cos
 \left( \theta \right)  \right) ^{3}- 0.03444\,{\it b_{pc}}\,{
{\rm e}^{- 1.0\,{\frac {r_{{{\it pc}}}\cos \left( \theta \right) }{{
\it b_{pc}}}}}}{r_{{{\it pc}}}}^{2} \left( \cos \left( \theta \right) 
 \right) ^{2}
\\
- 0.06888\,{{\it b_{pc}}}^{2}{{\rm e}^{- 1.0\,{\frac {
r_{{{\it pc}}}\cos \left( \theta \right) }{{\it b_{pc}}}}}}r_{{{\it pc}
}}\cos \left( \theta \right) - 0.06888\,{{\it b_{pc}}}^{3}{{\rm e}^
{- 1.0\,{\frac {r_{{{\it pc}}}\cos \left( \theta \right) }{{\it b_{pc}}}
}}}
\\
+ 0.03444\,{\it b_{pc}}\,{{\rm e}^{- 1.0\,{\frac {{\it r}_{{{
\it 0,pc}}}\cos \left( \theta \right) }{{\it b_{pc}}}}}}{{\it r}_{{{\it 
pc}}}}^{2} \left( \cos \left( \theta \right)  \right) ^{2}+
 0.06888\,{{\it b_{pc}}}^{2}{{\rm e}^{- 1.0\,{\frac {{\it r}_{{{
\it 0,pc}}}\cos \left( \theta \right) }{{\it b_{pc}}}}}}{\it r}_{{{\it 
0,pc}}}\cos \left( \theta \right) 
\\
+ 0.06888\,{{\it b_{pc}}}^{3}{
{\rm e}^{- 1.0\,{\frac {{\it r}_{{{\it 0,pc}}}\cos \left( \theta
 \right) }{{\it b_{pc}}}}}} \bigg ) n_{{0}}
\Bigg )
\Delta \Omega \,M_{\sun}
\quad ,
\end{multline}
where
$r_{0,pc}$ ,  $r_{pc}$ and  $b_{pc}$ 
are $r_0$  ,  $r$ and  $b$ expressed  in pc.

\subsection{A  Gaussian profile}

The density  is  assumed to have the following Gaussian
dependence on $z$
in Cartesian coordinates:
\begin{equation}
 \rho(z;b,\rho_0) =
\rho_0  {{\rm e}^{-\frac{1}{2}\,{\frac {{z}^{2}}{{b}^{2}}}}}
\quad ,
\label{profgaussian}
\end{equation}
where $b$ represents the standard deviation.
In spherical coordinates,
the density is
\begin{equation}
 \rho(r;r_0,b,\rho_0) =
  \begin{cases}
    \rho_0                                               & \quad \text{if } r    \leq r_0\\
    \rho_0 {{\rm e}^{-\frac{1}{2}\,{\frac {{z}^{2}}{{b}^{2}}}}}  & \quad \text{if } r    >   r_0\\
  \end{cases}
  \label{profgaussianr}
\quad .
\end{equation}

The total mass swept,   $M(r;r_0,b,\theta,\rho_0) $,
in the interval $[0,r]$
is
\begin{multline}
M (r;r_0,b,\theta,\rho_0)=
\Bigg (
\frac{1}{3}\,\rho_{{0}}{r_{{0}}}^{3}
+\rho_{{0}} \bigg( -{\frac {r{b}^{2}}{
 \left( \cos \left( \theta \right)  \right) ^{2}}{{\rm e}^{-\frac{1}{2}\,{
\frac {{r}^{2} \left( \cos \left( \theta \right)  \right) ^{2}}{{b}^{2
}}}}}}
\\
+\frac{1}{2}\,{\frac {{b}^{3}\sqrt {\pi}\sqrt {2}}{ \left( \cos \left( 
\theta \right)  \right) ^{3}}{\rm erf} \left(\frac{1}{2}\,{\frac {\sqrt {2}
\cos \left( \theta \right) r}{b}}\right)} \bigg ) 
-\rho_{{0}} \bigg( -
{\frac {r_{{0}}{b}^{2}}{ \left( \cos \left( \theta \right)  \right) ^{
2}}{{\rm e}^{-\frac{1}{2}\,{\frac {{r_{{0}}}^{2} \left( \cos \left( \theta
 \right)  \right) ^{2}}{{b}^{2}}}}}}
\\
+\frac{1}{2}\,{\frac {{b}^{3}\sqrt {\pi}
\sqrt {2}}{ \left( \cos \left( \theta \right)  \right) ^{3}}{\rm erf} 
\left(\frac{1}{2}\,{\frac {\sqrt {2}\cos \left( \theta \right) r_{{0}}}{b}}
\right)} \bigg ) 
\Bigg ) 
 \Delta \Omega
\quad ,
\label{massgaussian}
\end{multline}

\noindent where $\mathop{\mathrm{erf}}(x)$
is the error function, defined by
\begin{equation}
\mathop{\mathrm{erf}\/}\nolimits
(x)=\frac{2}{\sqrt{\pi}}\int_{0}^{x}e^{-t^{2}}dt
\quad .
\end{equation}

The previous formula expressed in solar masses is
\begin{multline}
M (r_{pc};r_{0,pc},b_{pc},\theta,n_0)= \\
\frac{1}{    ( \cos    ( \theta    )     ) ^{3}}
\Bigg (
- \bigg ( - 0.011481\,{{\it r}_{{{\it 0,pc}}}}^{3} \left( \cos
 \left( \theta \right)  \right) ^{3}+ 0.03444\,{{\rm e}^{- 0.5\,{
\frac {{r_{{{\it pc}}}}^{2} \left( \cos \left( \theta \right) 
 \right) ^{2}}{{{\it b_{pc}}}^{2}}}}}r_{{{\it pc}}}{{\it b_{pc}}}^{2}\cos
 \left( \theta \right)
\\
 - 0.04316\,{{\it b_{pc}}}^{3}{\rm erf} \left(
 0.7071\,{\frac {\cos \left( \theta \right) r_{{{\it pc}}}}{{
\it b_{pc}}}}\right)- 0.034443\,{{\rm e}^{- 0.5\,{\frac {{{\it r}_{{
{\it 0,pc}}}}^{2} \left( \cos \left( \theta \right)  \right) ^{2}}{{{
\it b_{pc}}}^{2}}}}}{\it r}_{{{\it 0,pc}}}{{\it b_{pc}}}^{2}\cos \left( 
\theta \right) 
\\
+ 0.04316\,{{\it b_{pc}}}^{3}{\rm erf} \left(
 0.7071\,{\frac {\cos \left( \theta \right) {\it r}_{{{\it 0,pc}
}}}{{\it b_{pc}}}}\right) \bigg ) n_{{0}}
\Bigg )
\Delta \Omega \,M_{\sun}
\quad ,
\end{multline}
where
$r_{0,pc}$ ,  $r_{pc}$ and  $b_{pc}$ 
are $r_0$  ,  $r$ and  $b$ expressed  in pc.

\section{The classical thin layer approximation}
\label{secclassical}

This section reviews  the standard
equation of motion in the case of
the thin layer approximation in the presence of 
a CSM  and derives the equation of motion  
for  the  density profiles here analysed.
One case of non-cubic dependence for the swept mass as a function of the
radius is also considered.

\subsection{Classical momentum conservation}

The conservation of the classical momentum in
spherical coordinates
along  the  solid angle  $\Delta \Omega$
in the framework of the thin
layer approximation  states that
\begin{equation}
M_0(r_0) \,v_0 = M(r)\,v
\quad ,
\end{equation}
where $M_0(r_0)$ and $M(r)$ are the swept masses at $r_0$ and $r$,
and $v_0$ and $v$ are the velocities of the thin layer at $r_0$ and $r$.
This conservation law can be expressed as a differential equation
of the first order by inserting $v=\frac{dr}{dt}$:
\begin{equation}
M(r)\, \frac{dr}{dt} - M_0\, v_0=0
\quad .
\end{equation}
A variant of the above equation introduces 
the non-cubic dependence (NCD), $p$,
which assumes that only a fraction
of the total mass enclosed in the  volume
of the expansion
accumulates  in a thin shell just after
the shock front.
The  global mass between  $0$ and $r_0$ 
along  the  solid angle  $\Delta \Omega$
is    $\frac {1}{3} \rho_0  r_0^3  $ where 
$\rho_0$ is the central density of the ambient medium.
The swept mass included in the  thin layer which characterizes the
expansion   is
\begin{equation}
M_0 =( \frac {1}{3}  \rho  r_0^3)^{\frac{1}{p}}
\quad  .
\end{equation}
Let us call $M(r)$ the   mass swept
between  $0$ and $r$ 
along  the  solid angle  $\Delta \Omega$.
The conservation of momentum in NCD gives
\begin{equation}
(M(r))  ^{\frac{1}{p}}    v  =
(M(r_0))^{\frac{1}{p}}  v_0 
\quad  ,
\quad NCD \, case
\label{conservationncd}
\end{equation}

The above two differential equations are independent
of the azimuthal angle $\varphi$.
The 3D surface which represents the advancing shock of the SN
is  generated by rotating 
the curve in the $x-z$ plane defined by
the analytical or numerical solution $r(t)$
about the $z$-axis and 
this is the {\it first symmetry}.
A  {\it second symmetry} around the $z=0$ plane
allows building the two lobes of the advancing surface.
The orientation  of the 3D surface  is characterized by
the Euler angles $(\Theta, \Phi, \Psi)$
and  therefore  by a total
$3 \times 3$  rotation matrix,
$E$, see \cite{Goldstein2002}.

The adopted astrophysical units  are pc
for length and
yr for time;
the initial velocity $v_{{0}}$ is
expressed in
pc yr$^{-1}$.
The astronomical velocities are evaluated in
km s$^{-1}$
and  therefore
$v_{{0}} =1.02\times10^{-6} v_{{1}}$
where  $v_{{1}}$ is the initial
velocity expressed in
km s$^{-1}$.

\subsection{Motion with an hyperbolic profile}

In the case of an hyperbolic density profile
for the CSM,
as given by equation (\ref{profhyperbolic}),
the differential equation
which models momentum conservation
is
\begin{equation}
 \left( \frac{1}{3}\, {r_{{0}}}^{3}+\frac{1}{2}\,{\frac { z_{{0}}
 \left( -{r_{{0}}}^{2}+ \left( r \left( t \right)  \right) ^{2}
 \right) }{\cos \left( \theta \right) }} \right) {\frac {\rm d}{
{\rm d}t}}r \left( t \right) -\frac{1}{3}\, {r_{{0}}}^{3}v_{{0}}=0
\quad ,
\end{equation}
where the initial  conditions
are  $r=r_0$  and   $v=v_0$
when $t=t_0$.
The variables can be separated and
the radius as a function of the time
is
\begin{equation}
r(t;t_0,z_0,v_0)= \frac{HN}{HD}
\quad ,
\nonumber
\label{rtanalyticalhyper}
\end{equation}
where
\begin{eqnarray}
HN= 
-\sqrt [3]{3} \Bigg  ( 2\,\cos   ( \theta   ) \sqrt [3]{3}r_{{0}}
-3\,\sqrt [3]{3}z_{{0}}- \bigg  ( -9\,{z_{{0}}}^{3/2}+   (    ( 9
\,t-9\,t_{{0}}   ) v_{{0}}
\nonumber  \\
+9\,r_{{0}}   ) \cos   ( \theta
   ) \sqrt {z_{{0}}}+\sqrt {3}\sqrt {27}\sqrt {{\it AHN}} \bigg  ) 
^{2/3}  \Bigg ) r_{{0}}
\quad ,
\end{eqnarray}
with
\begin{eqnarray}
AHN= 
 \Bigg ( {\frac {8\,   ( \cos   ( \theta   )    ) ^{2}{r_
{{0}}}^{3}}{27}}+z_{{0}}  \bigg (    ( t-t_{{0}}   ) ^{2}{v_{{0}}
}^{2}+2\,r_{{0}}   ( t-t_{{0}}   ) v_{{0}}-\frac{1}{3}\,{r_{{0}}}^{2}
  \bigg ) \cos   ( \theta   )
\nonumber \\
 -2\,v_{{0}}{z_{{0}}}^{2}   ( t-
t_{{0}}   )  \Bigg  ) \cos   ( \theta   ) 
\end{eqnarray}
and
\begin{eqnarray}
HD =  \nonumber \\
3\,\sqrt {z_{{0}}}\sqrt [3]{-9\,{z_{{0}}}^{3/2}+ \left(  \left( 9\,t-9
\,t_{{0}} \right) v_{{0}}+9\,r_{{0}} \right) \cos \left( \theta
 \right) \sqrt {z_{{0}}}+\sqrt {3}\sqrt {27}\sqrt {{\it BHD}}}
\quad ,  
\end{eqnarray}
with   
\begin{multline}
BHD=
\Bigg   ( {\frac {8\,   ( \cos   ( \theta   )    ) ^{2}{r_
{{0}}}^{3}}{27}}+z_{{0}} \bigg  (    ( t-t_{{0}}   ) ^{2}{v_{{0}}
}^{2}+2\,r_{{0}}   ( t-t_{{0}}   ) v_{{0}}-\frac{1}{3}\,{r_{{0}}}^{2}
\bigg   ) \cos   ( \theta   )
\nonumber  \\
 -2\,v_{{0}}{z_{{0}}}^{2}   ( t-
t_{{0}}   )   \Bigg  ) \cos   ( \theta   ) 
\quad .
\end{multline}
The velocity  as a function of the radius $r$  is
\begin{equation}
v(t)=
2\,{\frac {{r_{{0}}}^{3}v_{{0}}\cos \left( \theta \right) }{2\,{r_{{0}
}}^{3}\cos \left( \theta \right) -3\,{r_{{0}}}^{2}z_{{0}}+3\,{r}^{2}z_
{{0}}}}
\quad .
\end{equation}

\subsection{Motion with a power law profile}

In the case of a power-law density profile for the CSM
as given by equation~(\ref{profpower}),
the differential equation
which models the momentum conservation
is
\begin{multline}
-\frac{1}{3}\,{\frac {1}{\alpha-3} \Bigg ( 3\, \left( {\frac {z_{{0}}}{r_{{0}}
\cos \left( \theta \right) }} \right) ^{\alpha} \left( {\frac {\rm d}{
{\rm d}t}}r \left( t \right)  \right) {r_{{0}}}^{3}-3\, \left( r
 \left( t \right)  \right) ^{3} \left( {\frac {z_{{0}}}{r \left( t
 \right) \cos \left( \theta \right) }} \right) ^{\alpha}{\frac {\rm d}
{{\rm d}t}}r \left( t \right) 
}
\\
{
+ \left( {\frac {\rm d}{{\rm d}t}}r
 \left( t \right)  \right) {r_{{0}}}^{3}\alpha-{r_{{0}}}^{3}v_{{0}}
\alpha-3\, \left( {\frac {\rm d}{{\rm d}t}}r \left( t \right) 
 \right) {r_{{0}}}^{3}+3\,{r_{{0}}}^{3}v_{{0}} \Bigg) }=0
\label{eqndiffpower}
\quad .
\end{multline}
The velocity is
\begin{equation}
v(r;r_0,v_0,\theta,\alpha) =
\frac
{
-{r_{{0}}}^{3}v_{{0}} \left( \alpha-3 \right)
}
{
3\, \left( {\frac {z_{{0}}}{r\cos \left( \theta \right) }} \right) ^{
\alpha}{r}^{3}-3\, \left( {\frac {z_{{0}}}{r_{{0}}\cos \left( \theta
 \right) }} \right) ^{\alpha}{r_{{0}}}^{3}-{r_{{0}}}^{3}\alpha+3\,{r_{
{0}}}^{3}
}
\quad .
\end{equation}
We now evaluate the integral
\begin{equation}
I =\int  \frac{1}{v(r;r_0,v_0,\theta,\alpha)} dr
\quad ,
\end{equation}
which is
\begin{multline}
I(r) =3\,{\frac {{r}^{4}}{{r_{{0}}}^{3}v_{{0}} \left( \alpha-3 \right)
 \left( \alpha-4 \right) }{{\rm e}^{\alpha\,\ln  \left( {\frac {z_{{0}
}}{r\cos \left( \theta \right) }} \right) }}}
\\
+3\,{\frac {r}{v_{{0}}
 \left( \alpha-3 \right) } \left( {\frac {z_{{0}}}{r_{{0}}\cos \left(
\theta \right) }} \right) ^{\alpha}}+{\frac {\alpha\,r}{v_{{0}}
 \left( \alpha-3 \right) }}-3\,{\frac {r}{v_{{0}} \left( \alpha-3
 \right) }}
\end{multline}
The solution of the differential equation (\ref{eqndiffpower})
can  be found by solving numerically  the following nonlinear equation
\begin{equation}
I(r)- I(r_0)= t-t_0
\quad  .
\label{rtanalyticalpower}
\end{equation}

\subsection{Motion with an exponential profile}

In the case of  an exponential density profile
for the CSM
as given by equation (\ref{profexponential}),
the differential equation
which models momentum conservation
is
\begin{multline}
 \Bigg ( \frac{1}{3}\, {r_{{0}}}^{3}-{\frac {b \left(  \left( r
 \left( t \right)  \right) ^{2} \left( \cos \left( \theta \right) 
 \right) ^{2}+2\,r \left( t \right) b\cos \left( \theta \right) 
+2\,{b
}^{2} \right)  }{ \left( \cos \left( \theta \right)  \right) 
^{3}}{{\rm e}^{-{\frac {\cos \left( \theta \right) r \left( t \right) 
}{b}}}}}
\\
+{\frac {b \left( {r_{{0}}}^{2} \left( \cos \left( \theta
 \right)  \right) ^{2}+2\,r_{{0}}b\cos \left( \theta \right) 
+2\,{b}^{
2} \right)  }{ \left( \cos \left( \theta \right)  \right) ^{3
}}{{\rm e}^{-{\frac {r_{{0}}\cos \left( \theta \right) }{b}}}}}
 \Bigg ) {\frac {\rm d}{{\rm d}t}}r \left( t \right) =\frac{1}{3}\, {
r_{{0}}}^{3}v_{{0}}
\quad .
\label{eqndiffexp}
\end{multline}
There is no analytical solution.
We present the following
series solution of order 4 around $t_0$
\begin{multline}
r(t) =
r_{{0}}+ \left( t-{   t_0} \right) v_{{0}}-3/2\,{\frac {{v_{{0}}}^{2}
 \left( t-{   t_0} \right) ^{2}}{r_{{0}}}{{\rm e}^{-{\frac {r_{{0}}
\cos \left( \theta \right) }{b}}}}}
\\
+\frac{1}{2}\,{\frac {{v_{{0}}}^{3} \left(
t-{   t_0} \right) ^{3}}{{r_{{0}}}^{2}b}{{\rm e}^{-{\frac {r_{{0}}\cos
 \left( \theta \right) }{b}}}} \left( 9\,{{\rm e}^{-{\frac {r_{{0}}
\cos \left( \theta \right) }{b}}}}b+r_{{0}}\cos \left( \theta \right)
-2\,b \right) } 
\\
+ \orderof \left ( t-t_0 \right)^4
\label{rtseriesexp}
\quad .
\end{multline}
A second approximate solution can be found by
extracting  the velocity
from (\ref{eqndiffexp}):
\begin{equation}
v(r;r_0,v_0,\theta,b) =\frac{VN}{VD}
\end{equation}
where
\begin{equation}
VN= -{r_{{0}}}^{3}v_{{0}} \left( \cos \left( \theta \right)  \right) ^{3}
\quad ,
\end{equation}
and
\begin{multline}
VD=
3\,{{\rm e}^{-{\frac {r\cos   ( \theta   ) }{b}}}}   ( \cos
   ( \theta   )    ) ^{2}b{r}^{2}-3\,{{\rm e}^{-{\frac {r_{
{0}}\cos   ( \theta   ) }{b}}}}   ( \cos   ( \theta
   )    ) ^{2}{r_{{0}}}^{2}b-   ( \cos   ( \theta
   )    ) ^{3}{r_{{0}}}^{3}
   \\
   +6\,{{\rm e}^{-{\frac {r\cos
   ( \theta   ) }{b}}}}\cos   ( \theta   ) {b}^{2}r
   -6\,{
{\rm e}^{-{\frac {r_{{0}}\cos   ( \theta   ) }{b}}}}\cos
   ( \theta   ) r_{{0}}{b}^{2}
  \\
    +6\,{{\rm e}^{-{\frac {r\cos
   ( \theta   ) }{b}}}}{b}^{3}-6\,{{\rm e}^{-{\frac {r_{{0}}
\cos   ( \theta   ) }{b}}}}{b}^{3}
\quad .
\end{multline}
Given a function $f(r)$, the Pad\'e  approximant,
after \cite{Pade1892},
is
\begin{equation}
f(r)=\frac{a_{0}+a_{1}r+\dots+a_{p}r^{o}}{b_{0}+b_{1}%
r+\dots+b_{q}r^{q}}
\quad ,
\end{equation}
where the notation is the same as in \cite{NIST2010}.
The coefficients $a_i$ and $b_i$
are found through Wynn's cross rule,
see \cite{Baker1975,Baker1996},
and our choice is $o=2$ and $q=1$.
The choice of  $o$ and $q$ is a compromise between
precision, needing high values for  $o$ and $q$, and
simplicity of the expressions to manage,
needing low values for  $o$ and $q$.
The inverse of the velocity expressed by the
the Pad\`e approximant
is
\begin{equation}
(\frac{1}{v(r)})_{2,1} = \frac{N21}{D21}
\quad .
 \end{equation}
 where 
\begin{multline}
N21=   ( r-r_{{0}}   )    ( 9\,{{\rm e}^{-{\frac {r_{{0}}\cos
   ( \theta   ) }{b}}}}br-9\,{{\rm e}^{-{\frac {r_{{0}}\cos
   ( \theta   ) }{b}}}}br_{{0}}
   \\+2\,\cos   ( \theta   ) r
r_{{0}}-2\,\cos   ( \theta   ) {r_{{0}}}^{2}-4\,br+10\,r_{{0}}b
   )
\end{multline}
and
\begin{equation}
D21=
2\,v_{{0}} \left( \cos \left( \theta \right) rr_{{0}}-\cos \left(
\theta \right) {r_{{0}}}^{2}-2\,br+5\,r_{{0}}b \right)
\end{equation}
 
The above result allows deducing a solution $r_{2,1}$
expressed through the Pad\`e approximant
\begin{equation}
r(t)_{2,1} =
\frac
{
B+\sqrt {A}
}
{
9\,{{\rm e}^{-{\frac {r_{{0}}\cos \left( \theta \right) }{b}}}}b+2\,r_
{{0}}\cos \left( \theta \right) -4\,b
}
\label{rtpade}
\end{equation}
where
\begin{multline}
A=
 \left( \cos \left( \theta \right)  \right) ^{2}{r_{{0}}}^{2}{t}^{2}{v
_{{0}}}^{2}-2\, \left( \cos \left( \theta \right)  \right) ^{2}{r_{{0}
}}^{2}tt_{{0}}{v_{{0}}}^{2}+ \left( \cos \left( \theta \right)
 \right) ^{2}{r_{{0}}}^{2}{t_{{0}}}^{2}{v_{{0}}}^{2}
\\
-4\,\cos \left(
\theta \right) r_{{0}}b{t}^{2}{v_{{0}}}^{2}+8\,\cos \left( \theta
 \right) r_{{0}}btt_{{0}}{v_{{0}}}^{2}
-4\,\cos \left( \theta \right) r
_{{0}}b{t_{{0}}}^{2}{v_{{0}}}^{2}
\\
+54\,{{\rm e}^{-{\frac {r_{{0}}\cos
 \left( \theta \right) }{b}}}}r_{{0}}{b}^{2}tv_{{0}}-54\,{{\rm e}^{-{
\frac {r_{{0}}\cos \left( \theta \right) }{b}}}}r_{{0}}{b}^{2}t_{{0}}v
_{{0}}+6\,\cos \left( \theta \right) {r_{{0}}}^{2}btv_{{0}}
\\
-6\,\cos
 \left( \theta \right) {r_{{0}}}^{2}bt_{{0}}v_{{0}}+4\,{b}^{2}{t}^{2}{
v_{{0}}}^{2}-8\,{b}^{2}tt_{{0}}{v_{{0}}}^{2}
+4\,{b}^{2}{t_{{0}}}^{2}{v
_{{0}}}^{2}-12\,r_{{0}}{b}^{2}tv_{{0}}
\\
+12\,r_{{0}}{b}^{2}t_{{0}}v_{{0}
}+9\,{b}^{2}{r_{{0}}}^{2}
\end{multline}
and
\begin{multline}
B=
r_{{0}}tv_{{0}}\cos \left( \theta \right) -r_{{0}}t_{{0}}v_{{0}}\cos
 \left( \theta \right) +9\,{{\rm e}^{-{\frac {r_{{0}}\cos \left(
\theta \right) }{b}}}}br_{{0}}+2\,\cos \left( \theta \right) {r_{{0}}}
^{2}
\\
-2\,btv_{{0}}+2\,bt_{{0}}v_{{0}}-7\,r_{{0}}b
\quad .
\end{multline}

\subsection{Motion with a Gaussian profile}

In the case of  a Gaussian density  profile
for the CSM
as given by equation (\ref{profgaussian}),
the differential equation
which models momentum conservation
is
\begin{multline}
\Bigg (
 \frac{1}{3}\, {r_{{0}}}^{3}+  \Bigg ( -{\frac {r
 \left( t \right) {b}^{2}}{ \left( \cos \left( \theta \right) 
 \right) ^{2}}{{\rm e}^{-\frac{1}{2}\,{\frac { \left( r \left( t \right) 
 \right) ^{2} \left( \cos \left( \theta \right)  \right) ^{2}}{{b}^{2}
}}}}}
\\
+\frac{1}{2}\,{\frac {{b}^{3}\sqrt {\pi}\sqrt {2}}{ \left( \cos \left( 
\theta \right)  \right) ^{3}}{\rm erf} \left(\frac{1}{2}\,{\frac {\sqrt {2}
\cos \left( \theta \right) r \left( t \right) }{b}}\right)} \Bigg ) 
\\
-
  \left( -{\frac {r_{{0}}{b}^{2}}{ \left( \cos \left( \theta
 \right)  \right) ^{2}}{{\rm e}^{-\frac{1}{2}\,{\frac {{r_{{0}}}^{2} \left( 
\cos \left( \theta \right)  \right) ^{2}}{{b}^{2}}}}}}+\frac{1}{2}\,{\frac {{b
}^{3}\sqrt {\pi}\sqrt {2}}{ \left( \cos \left( \theta \right) 
 \right) ^{3}}{\rm erf} \left(\frac{1}{2}\,{\frac {\sqrt {2}\cos \left( \theta
 \right) r_{{0}}}{b}}\right)} \right)  
\Bigg ) 
{\frac {\rm d}{{\rm d}t
}}r \left( t \right) 
\\
=\frac{1}{3}\, {r_{{0}}}^{3}v_{{0}}
\quad  .
\label{eqndiffgauss}
\end{multline}
There is no analytical solution.

\subsection{Motion with an exponential profile, NCD }

In the case of  an exponential density profile
for the CSM
as given by equation (\ref{profexponential}),
the differential equation
which models momentum conservation
in the NCD case, see equation (\ref{conservationncd}),
is  
\begin{equation}
-{3}^{-{p}^{-1}}{r_{{0}}}^{3\,{p}^{-1}}v_{{0}}+ \left( {\frac {\rm d}{
{\rm d}t}}r \left( t \right)  \right) 
\biggl 
({{\frac{1}{3}\,{\frac {{   ANCD
}}{ \left( \cos \left( \theta \right)  \right) ^{3}}}}}
\biggr)^{\frac{1}{p}}=0 
\quad ,
\quad NCD \, case
\end{equation}
with
\begin{eqnarray}
ANCD=
\left( - \left( r \left( t \right)  \right) ^{2} \left( \cos \left( 
\theta \right)  \right) ^{2}b-2\,r \left( t \right) \cos \left( \theta
 \right) {b}^{2}-2\,{b}^{3} \right) {{\rm e}^{-{\frac {r \left( t
 \right) \cos \left( \theta \right) }{b}}}}
\\
+ \left( b{r_{{0}}}^{2}
 \left( \cos \left( \theta \right)  \right) ^{2}+2\,{b}^{2}r_{{0}}\cos
 \left( \theta \right) +2\,{b}^{3} \right) {{\rm e}^{-{\frac {r_{{0}}
\cos \left( \theta \right) }{b}}}}+{r_{{0}}}^{3} \left( \cos \left( 
\theta \right)  \right) ^{3}
\quad .
\label{eqndiffexpncd}
\end{eqnarray}
The above  differential equation does not have
an analytical solution and should be solved numerically.

\section{The relativistic thin layer approximation}
\label{secrelativistic}

This section reviews the conservation 
of momentum in special relativity 
and subsequently introduces two laws of motion.

\subsection{Relativistic momentum conservation}

The conservation of relativistic momentum
in spherical coordinates
along  the  solid angle  $\Delta \Omega$
in the framework of the thin
layer approximation  gives
\begin{equation}
{\frac {M \left( t ;\theta\right) \beta}{\sqrt {1-{\beta}^{2}}}}
\Delta \Omega
={\frac {M_{
{0}} \left( t_{{0}} \right) \beta0}{\sqrt {1-{\beta0}^{2}}}}
\Delta \Omega
\quad .
\end{equation}
Here, $M_0(r_0)$ and $M(r)$ are the swept
masses at $r_0$ and $r$,
$\beta=\frac{v}{c}$, $\beta_0=\frac{v_0}{c}$,
and $v_0$ and $v$ are the velocities of the thin
layer at $r_0$ and $r$,
see \cite{Zaninetti2014c,Zaninetti2016d} for more details.
We have chosen as units pc for distances and
yr for time, and therefore   the speed of light is
$c=0.306$\ pc\ yr$^{-1}$.

\subsection{Relativistic motion with a hyperbolic profile}

In the case of a hyperbolic density profile for the CSM
as given by equation (\ref{profhyperbolic}),
the differential equation
which models the relativistic momentum conservation
is
\begin{equation}
\frac
{
   \left( 2\,{r_{{0}}}^{3}\cos \left( \theta \right) +3\,{\it 
z0}\, \left( r \left( t \right)  \right) ^{2}-3\,{\it z0}\,{r_{{0}}}^{
2} \right) {\frac {\rm d}{{\rm d}t}}r \left( t \right) 
}
{
\,\cos \left( \theta \right) c\sqrt {1-{\frac { \left( {\frac {\rm d}{
{\rm d}t}}r \left( t \right)  \right) ^{2}}{{c}^{2}}}}
}
-
\frac
{
  {r_{{0}}}^{3}\beta_{{0}}
}
{
3\,\sqrt {1 -{\beta_{{0}}}^{2}}
}
=0
\label{eqndiffrelhyper}
\quad  ,
\end{equation}
where the initial  conditions
are  $r=r_0$  and   $v=v_0$
when $t=t_0$.

The velocity expressed in terms of $\beta$ can be derived
from the above equation:
\begin{equation}
\beta = \frac
{
2\,{r_{{0}}}^{3}\cos \left( \theta \right) \beta_{{0}}
}
{
D
}
\quad ,
\end{equation}
where
\begin{multline}
D=
\Bigl (
12\,{r_{{0}}}^{5}{\beta_{{0}}}^{2}{   z0}\,\cos \left( \theta
 \right) -12\,{r_{{0}}}^{3}{\beta_{{0}}}^{2}{r}^{2}{   z0}\,\cos
 \left( \theta \right) 
\\
+4\,{r_{{0}}}^{6} \left( \cos \left( \theta
 \right)  \right) ^{2}-9\,{r_{{0}}}^{4}{\beta_{{0}}}^{2}{{   z0}}^{2} \\
+18\,{r_{{0}}}^{2}{\beta_{{0}}}^{2}{r}^{2}{{   z0}}^{2}-9\,{\beta_{{0
}}}^{2}{r}^{4}{{   z0}}^{2}-12\,{r_{{0}}}^{5}{   z0}\,\cos \left(
\theta \right) 
\\
+12\,{r_{{0}}}^{3}{r}^{2}{   z0}\,\cos \left( \theta
 \right)
   +9\,{r_{{0}}}^{4}{{   z0}}^{2}-18\,{r_{{0}}}^{2}{r}^{2}{{
   z0}}^{2}+9\,{r}^{4}{{   z0}}^{2} \Bigr )^{\frac{1}{2}}
\quad .
\end{multline}
There is no analytical solution
of (\ref{eqndiffrelhyper}),
but we present the following
series solution of order three around $t_0$:
\begin{equation}
r(t)=r_{{0}}+\beta_{{0}}c \left( t-{   t0} \right) +\frac{3}{2}  {\frac {{c}^{2}
 \left( {\beta_{{0}}}^{2}-1 \right) z_{{0}}{\beta_{{0}}}^{2} \left( t-
{   t0} \right) ^{2}}{{r_{{0}}}^{2}\cos \left( \theta \right) }}
+ \orderof \left ( t-t_0 \right)^3
\label{rtserieshyperrel}
\quad .
\end{equation}
More details can  be found in \cite{Zaninetti2012b}.

\subsection{Relativistic motion with an exponential profile}

In the case of an exponential density profile for the CSM
as given by equation (\ref{profexponential}),
the differential equation
which models the relativistic momentum conservation
is
\begin{equation}
\frac
{
  N{\frac {\rm d}{{\rm d}t}}r \left( t \right)
}
{
3\, \left( \cos \left( \theta \right)  \right) ^{3}c\sqrt {1-{\frac {
 \left( {\frac {\rm d}{{\rm d}t}}r \left( t \right)  \right) ^{2}}{{c}
^{2}}}}
}
-
\frac
{
  {r_{{0}}}^{3}\beta_{{0}}
}
{
3\,\sqrt {1-{\beta_{{0}}}^{2}}
}
=0
\quad  ,
\end{equation}
where
\begin{multline}
N=-
3\,{{\rm e}^{-{\frac {r_{{0}}\cos \left( \theta \right) }{b}}}}
 \left( \cos \left( \theta \right)  \right) ^{2}{r_{{0}}}^{2}b-3\,
 \left( r \left( t \right)  \right) ^{2}{{\rm e}^{-{\frac {r \left( t
 \right) \cos \left( \theta \right) }{b}}}} \left( \cos \left( \theta
 \right)  \right) ^{2}b+ \left( \cos \left( \theta \right)  \right) ^{
3}{r_{{0}}}^{3}
\\
+6\,{{\rm e}^{-{\frac {r_{{0}}\cos \left( \theta
 \right) }{b}}}}\cos \left( \theta \right) r_{{0}}{b}^{2}-6\,r \left( 
t \right) {{\rm e}^{-{\frac {r \left( t \right) \cos \left( \theta
 \right) }{b}}}}\cos \left( \theta \right) {b}^{2}+6\,{{\rm e}^{-{
\frac {r_{{0}}\cos \left( \theta \right) }{b}}}}{b}^{3}
\\
-6\,{{\rm e}^{-
{\frac {r \left( t \right) \cos \left( \theta \right) }{b}}}}{b}^{3}
   \label{eqndiffrelexp}
\quad  .
\end{multline}
There is no analytical solution
of (\ref{eqndiffrelexp}),
so we present the following
series solution of order three around $t_0$:
\begin{multline}
r(t)=
\\
r_{{0}}+c\beta_{{0}} \left( t-t_{{0}} \right) +\frac{3}{2}\,{\frac {{\beta_{{0
}}}^{2}{c}^{2} \left( {\beta_{{0}}}^{2}-1 \right)  \left( t-t_{{0}}
 \right) ^{2}}{r_{{0}}}{{\rm e}^{-{\frac {r_{{0}}\cos \left( \theta
 \right) }{b}}}}}
+ \orderof \left ( t-t_0 \right)^3
\label{rtseriesexprrel}
\quad  .
\end{multline}

\section{The simulated objects}

This section reviews the astronomical observations
of \sn1987a and \s1006.

\subsection{Great asymmetry,  \sn1987a}

\begin{figure*}
\begin{center}
\includegraphics[width=7cm]{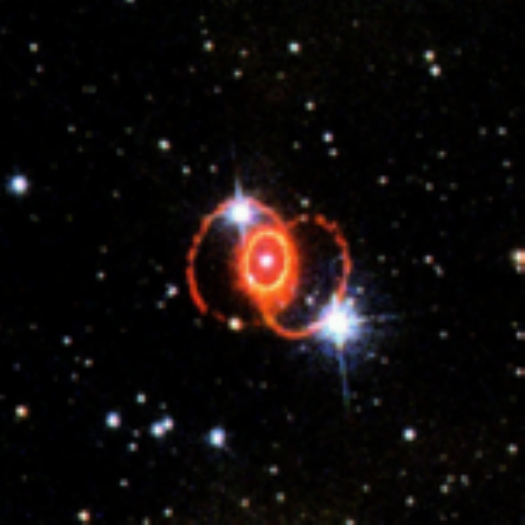}
\end {center}
\caption
{
An ST image of \sn1987a in the year 1997.   
}
\label{sn1987a_st}
    \end{figure*}

The complex structure of \sn1987a,
see an Hubble Space Telescope (ST) image in Figure
\ref{sn1987a_st},
can be classified as
a torus only, a torus plus two lobes, and a torus plus 4 lobes,
see \cite{Racusin2009,McCray2017}.

\begin{figure*}
\begin{center}
\includegraphics[width=7cm]{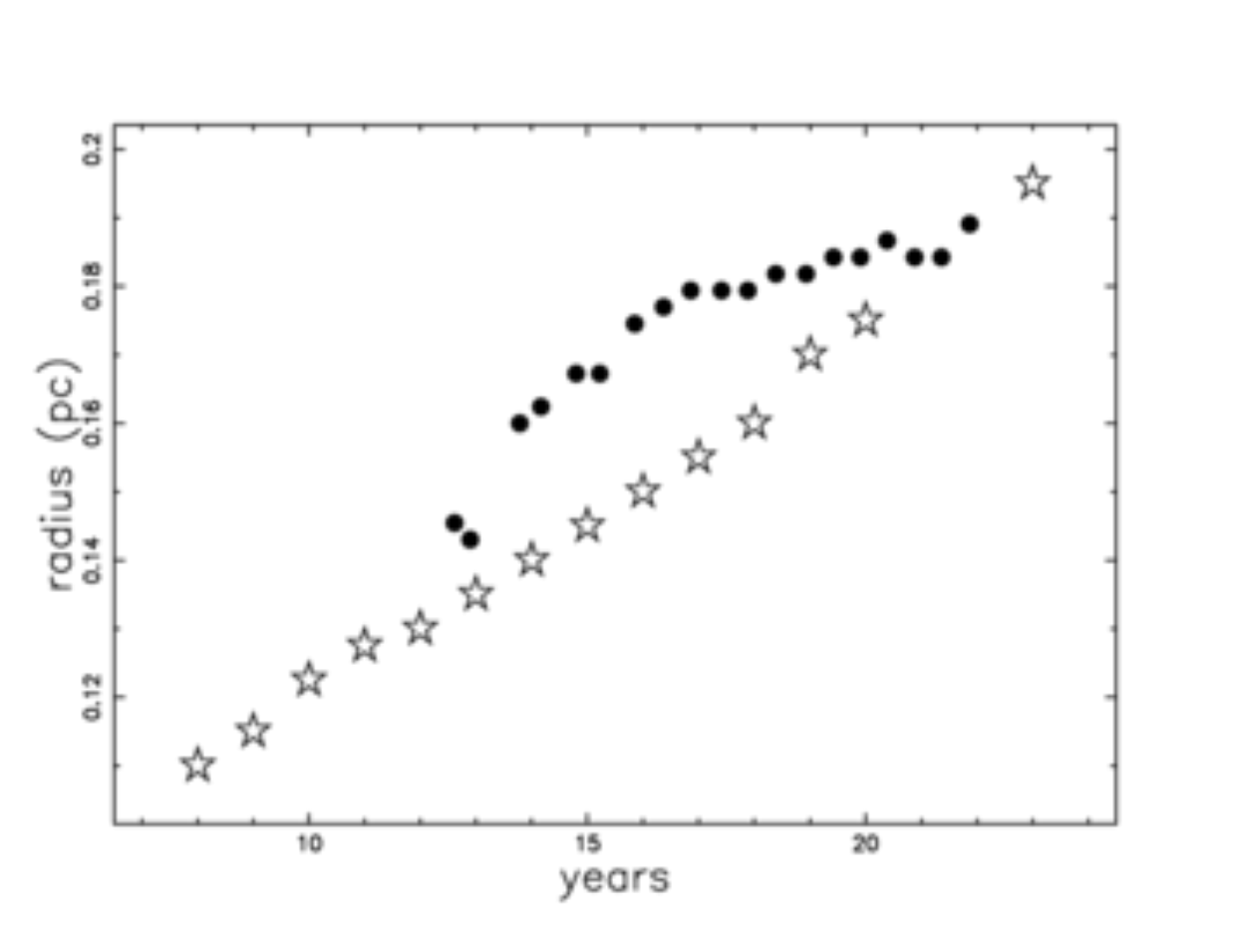}
\end {center}
\caption
{
Observed  radius  for the  torus only  
of \sn1987a
as function of time:
full points as  in Racusin et al. 2009
  and
  empty stars as  in Chiad et al.  2012.
}
\label{twodataeq}
    \end{figure*}

\begin{figure*}
\begin{center}
\includegraphics[width=7cm]{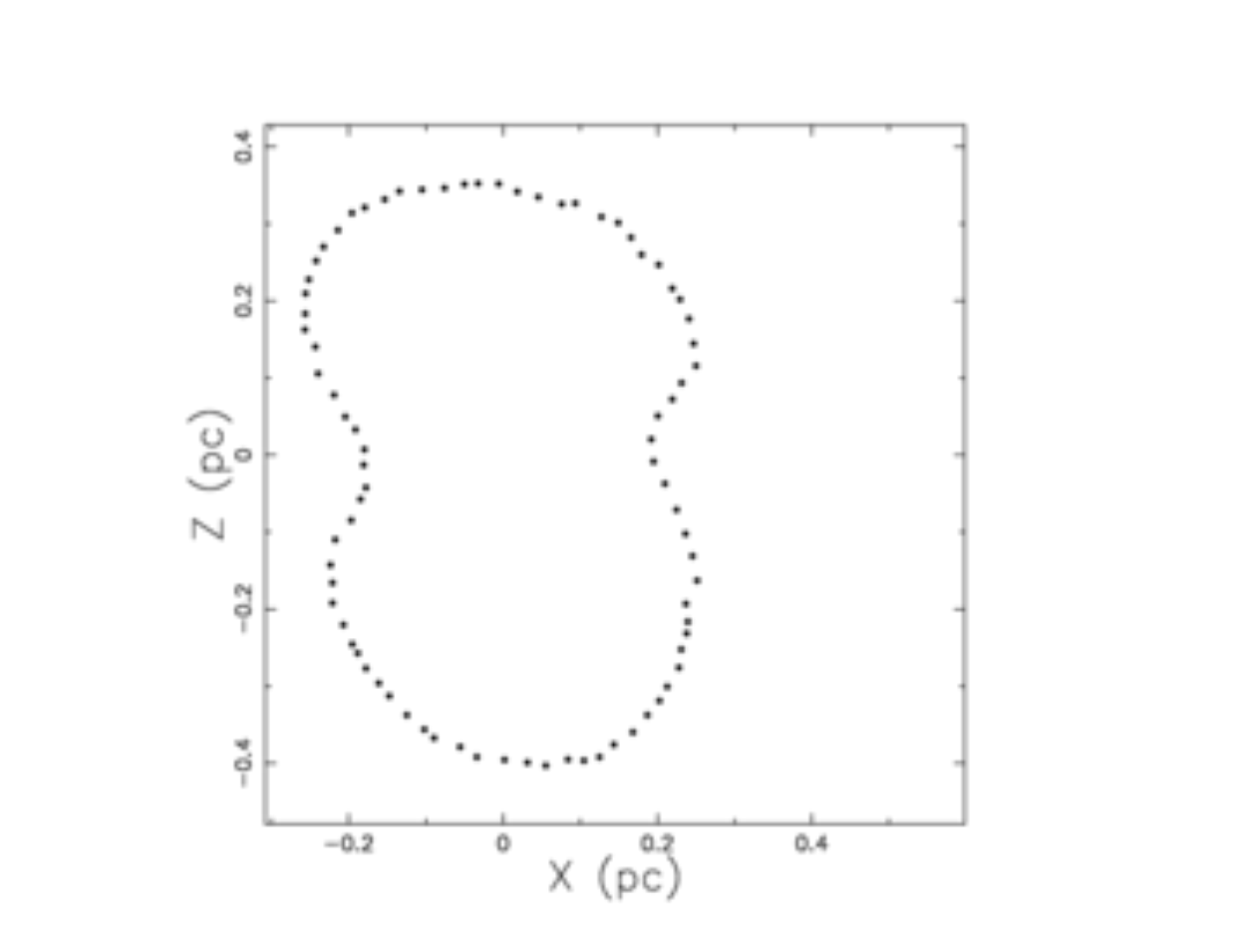}
\end {center}
\caption
{
Geometrical section of \sn1987a
in the $x-z$ plane adapted by the author from Fig.  5
in \cite{France2015}.
}
\label{section_obs_sn1987a}
    \end{figure*}

The region connected with the
radius of the advancing torus is here identified
with  our equatorial
region,
in  spherical coordinates, $\theta =\frac{\pi}{2}$.
The radius of the torus only  as a function of time can be found in
Table 2 of \cite{Racusin2009} or Fig.  3 of \cite{Chiad2012},
see Figure  \ref{twodataeq} for a comparison of the
two different techniques.
The radius of the torus only as given by the
method  of counting pixels (\cite{Chiad2012})
shows a more regular behaviour
and we have calibrated our codes in the equatorial region
in such a way that at time
23 years   the radius is $\frac{0.39}{2}$\,pc.
Another useful resource for calibration is the geometrical  section
of \sn1987a
reported as a sketch in Fig.  5 of \cite{France2015}.
This geometrical  section
was digitized  and  rotated in the $x-z$ plane
by $-40\degreezan$,  see Figure \ref{section_obs_sn1987a}.

An hydrodynamical simulation of the torus only for  \sn1987a 
is reported in Figure 4 of \cite{Orlando2015}.  

\subsection{Weak  asymmetry,   \s1006 }

\begin{figure*}
\begin{center}
\includegraphics[width=7cm]{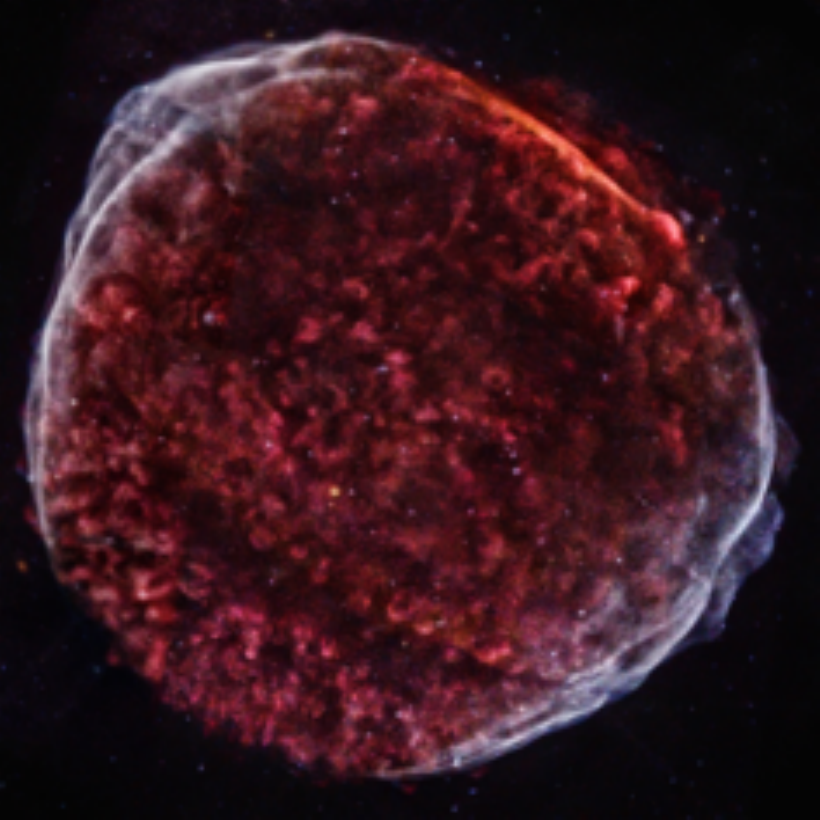}
\end {center}
\caption
{
A Chandra  X-ray (Red, Green, Blue)  image of \s1006 in the year 2013.   
}
\label{SN_CHANDRA_X_2013}
    \end{figure*}

\begin{figure*}
\begin{center}
\includegraphics[width=7cm]{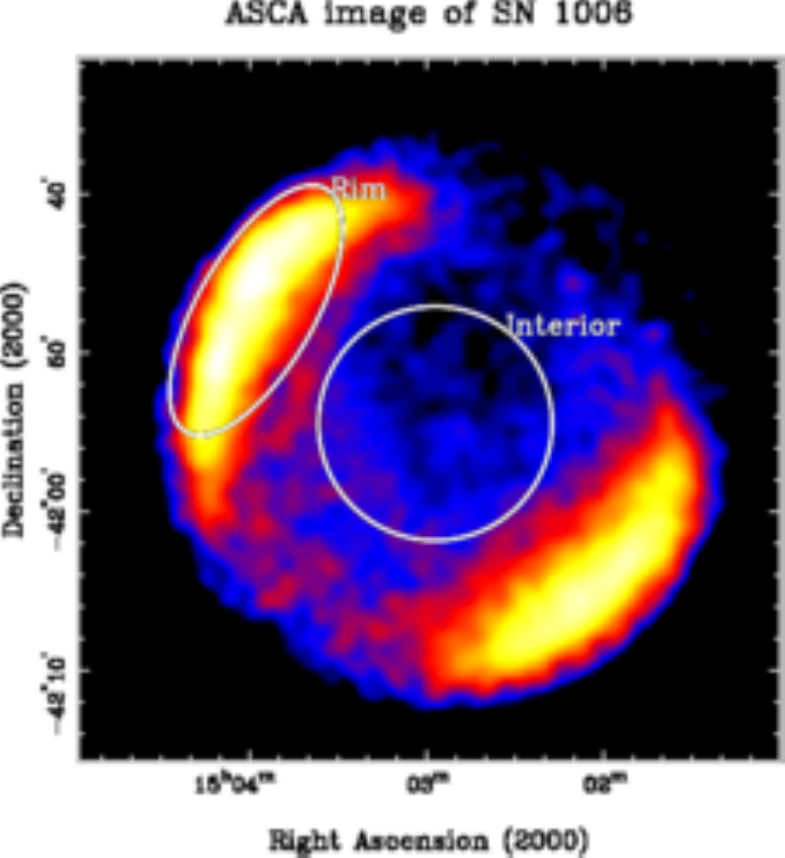}
\end {center}
\caption
{
An Advanced Satellite for Cosmology and Astrophysics (ASCA)  X-ray image of \s1006.   
}
\label{sn1006asca}
    \end{figure*}

This SN  started to be
visible in 1006 AD
and currently has a
radius  of 12.2 \mbox{pc}, see \cite{Uchida2013,Katsuda2017}.
The X-shape can be visualized in Figure \ref{SN_CHANDRA_X_2013}
and Figure \ref{sn1006asca}.

\begin{figure*}
\begin{center}
\includegraphics[width=7cm]{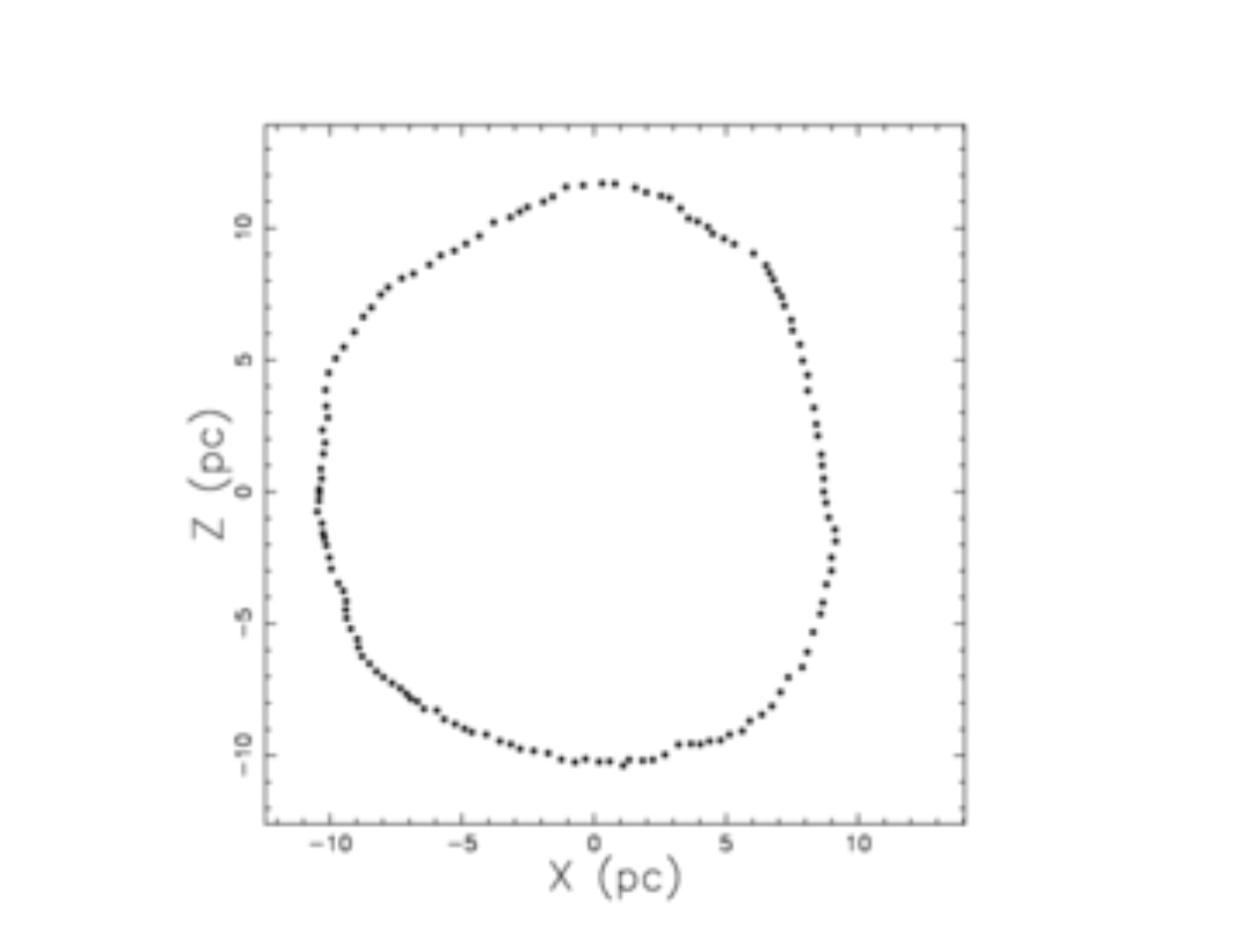}
\end {center}
\caption
{
Geometrical section of \s1006
in the $x-z$ plane adapted by the author from 
our Figure \ref{SN_CHANDRA_X_2013}. 
}
\label{sn1006_obs}
    \end{figure*}

More precisely, on
referring to the 
above X-image, 
it   can be observed that the radius is
greatest in the north-east direction,
see also the radio map of \s1006  at 1370 \mbox{MHz} by
\cite {Reynolds1986},
and the X-map in the 0.4–5.0 keV band 
of Fig.  1 in \cite{Uchida2013}.
The following observed radii can be extracted:
$R_{up}=11.69$~pc in the polar  direction and $R_{eq}= 8.7$~pc in the
equatorial direction.
A geometrical section of the above X-map
was digitized  and  rotated in the $x-z$ plane
by $-45\degreezan$,  see Figure \ref{sn1006_obs}.

Many filament radial intensity profiles at 2–7 keV
for \s1006  can be found in Fig. 8 of  \cite{Ressler2014}.
According to  \cite{Uchida2013},
the velocity of the ejecta should  be $\approx 3100 \frac{km}{s}$.

\section{Astrophysical  Results}

This section reports  the theoretical results compared 
with the astronomical observations 
for   \sn1987a and \s1006.

\subsection{The case \sn1987a}

\begin{figure*}
\begin{center}
\includegraphics[width=7cm]{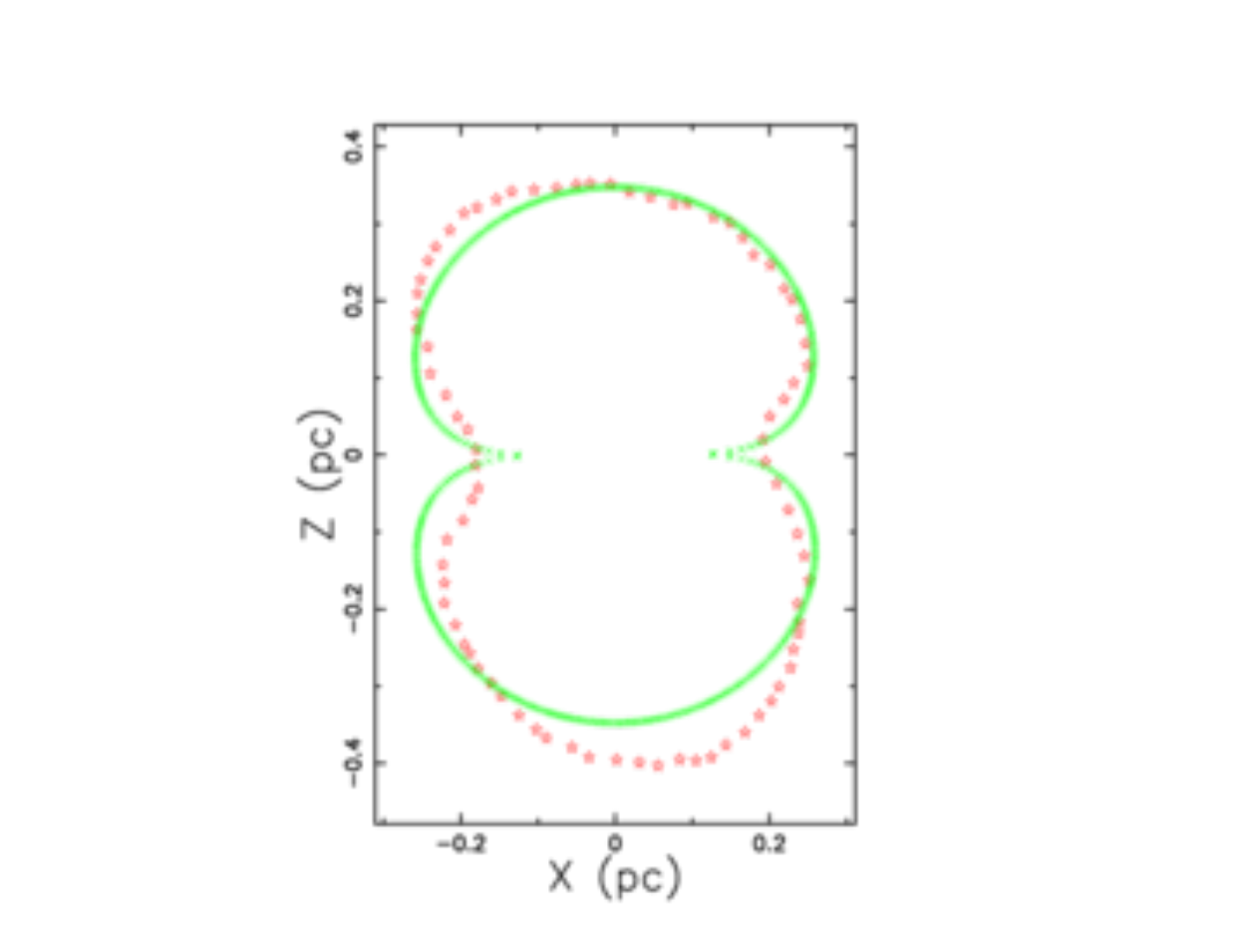}
\end {center}
\caption
{
Geometrical section of \sn1987a
in the $x-z$ plane with an hyperbolic profile
(green points)
and the observed profile (red stars).
The parameters
$r_0=0.06$\ pc,
$z_0=0.001$\ pc,
$t=21.86$\ yr,
$t_0 =0.1$ yr  and
$v_0\,=25000$ km s$^{-1}$
give
$\epsilon_{\mathrm {obs}}=92.13\%$.
}
\label{cut_hyper_1987a}
    \end{figure*}

\begin{figure*}
\begin{center}
\includegraphics[width=7cm]{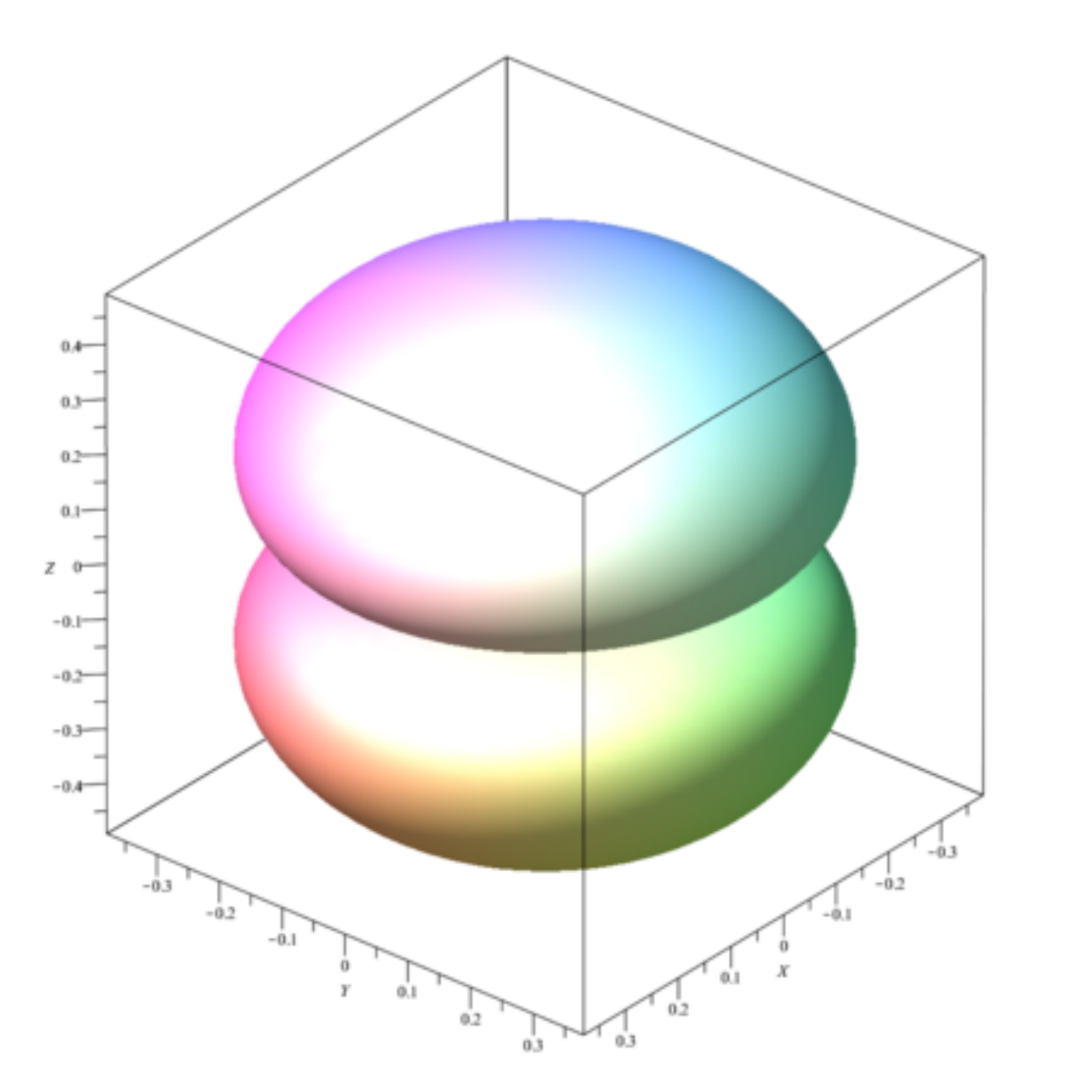}
\end {center}
\caption
{
3D surface  of  \sn1987a
with parameters as in Figure \ref{cut_hyper_1987a}, 
hyperbolic profile.
The three Euler angles are $\Theta=40$, $\Phi=60$ and
$ \Psi=60 $.
}
\label{3dsurfacehyper}
    \end{figure*}

In the case  of an {\it hyperbolic}
 density profile,
see  equation (\ref{profhyperbolicr}),
we have an analytical solution for the
motion, see equation (\ref{rtanalyticalhyper}).
Figure \ref{cut_hyper_1987a} displays a cut  of  \sn1987a
in the $x-z$ plane for the hyperbolic case.
For the hyperbolic case 
the observational
percentage reliability, 
see formula (\ref{efficiencymany}),
is $\epsilon_{\mathrm {obs}}=92.13\%$.
A rotation around  the
$z$-axis  of the previous  geometrical section allows
building a 3D surface, see
Figure \ref{3dsurfacehyper}.

\begin{figure*}
\begin{center}
\includegraphics[width=7cm]{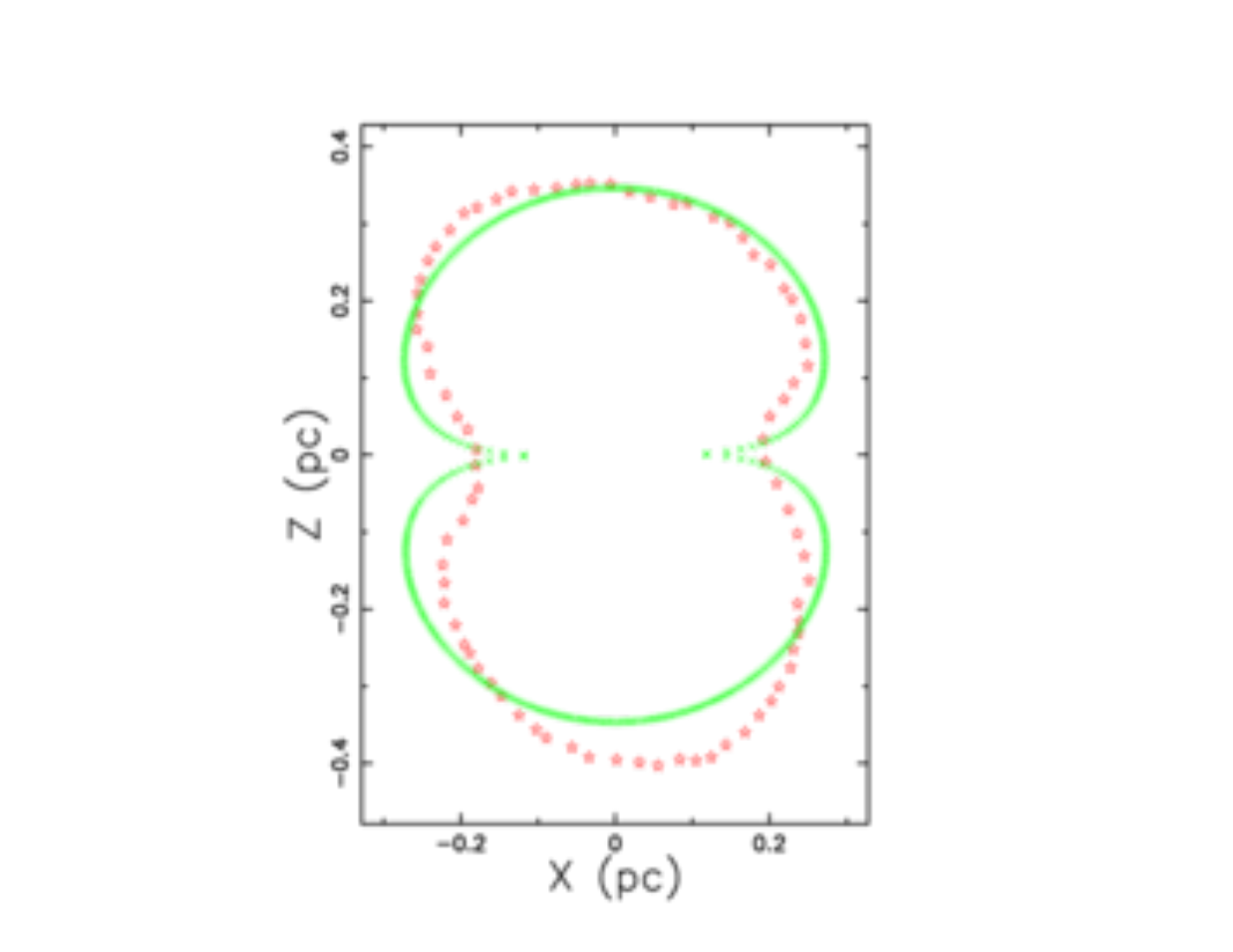}
\end {center}
\caption
{
Geometrical section of \sn1987a
in the $x-z$ plane with a power law profile
(green points)
and observed profile (red stars).
The parameters
$r_0=0.05$ pc,
$z_0=0.002$\ pc,
$t=21.86$\ yr,
$t_0 =0.1$\ yr,
$\alpha=1.3$,
$v_0\,=15000$ \ km s$^{-1}$
give
$\epsilon_{\mathrm {obs}}=90.76\%$.
}
\label{cut_power_1987a}
    \end{figure*}

In the case  of a {\it power law }
density profile,
see  equation (\ref{profpower}),
we evaluate a   numerical  solution for the
motion, see the nonlinear equation (\ref{rtanalyticalpower}).
Figure \ref{cut_power_1987a} reports  the numerical
solution for the power law case as a  cut  of  \sn1987a
in the $x-z$ plane.

\begin{figure*}
\begin{center}
\includegraphics[width=7cm]{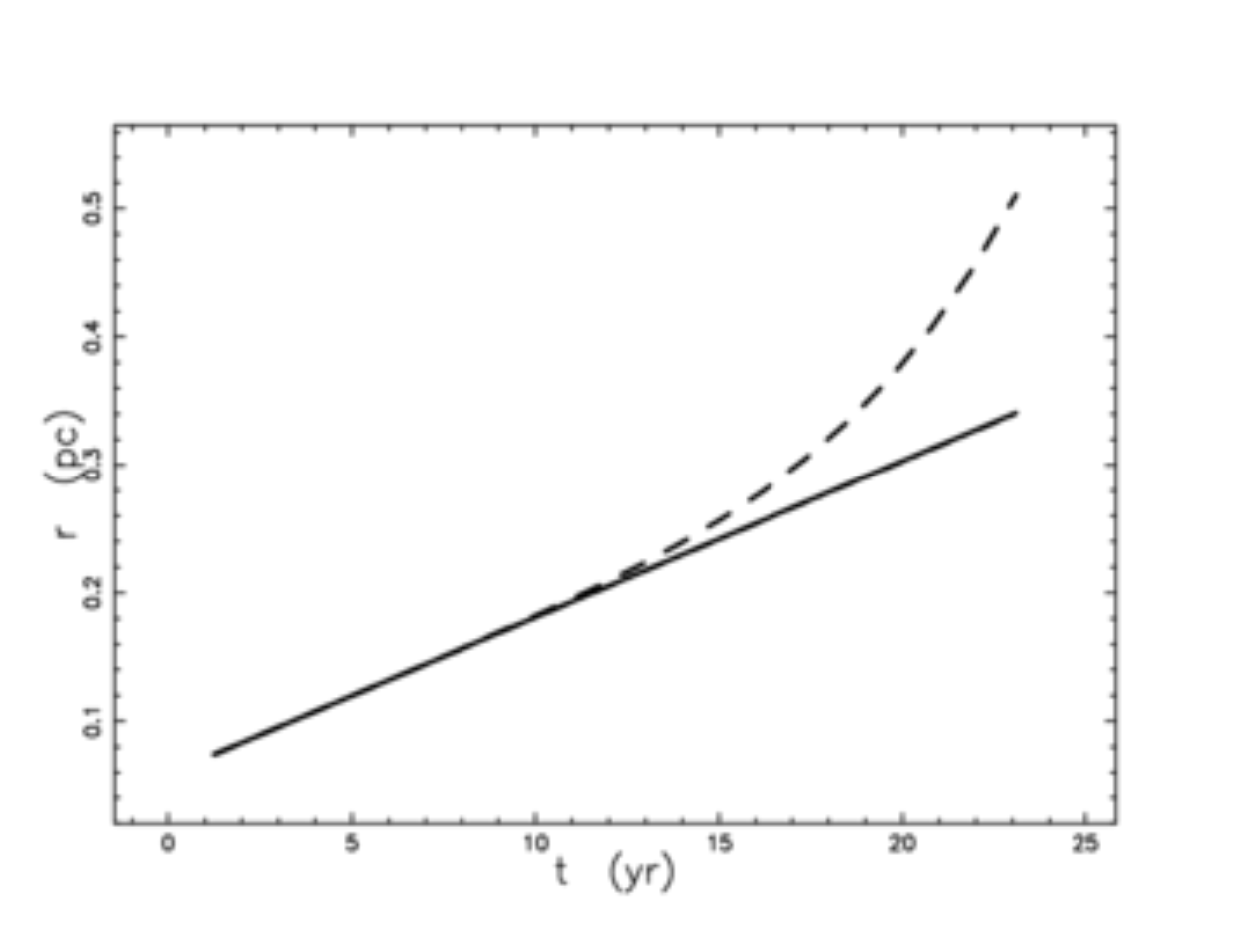}
\end {center}
\caption
{
Numerical   solution         (full   line),
power series solution        (dashed line)
and
Pad\'e  approximant solution (dot-dash-dot-dash line)
which is nearly equal to the numerical solution;
exponential profile.
The parameters are
$r_0=0.06$ pc,
$t=21.86$\ yr,
$t_0 =0.1$\ yr,
$b= 0.011$ pc,
$\theta=0$,
and
$v_0\,=12000 \ km s^{-1}$.
}
\label{pade_exp}
    \end{figure*}

\begin{figure*}
\begin{center}
\includegraphics[width=7cm]{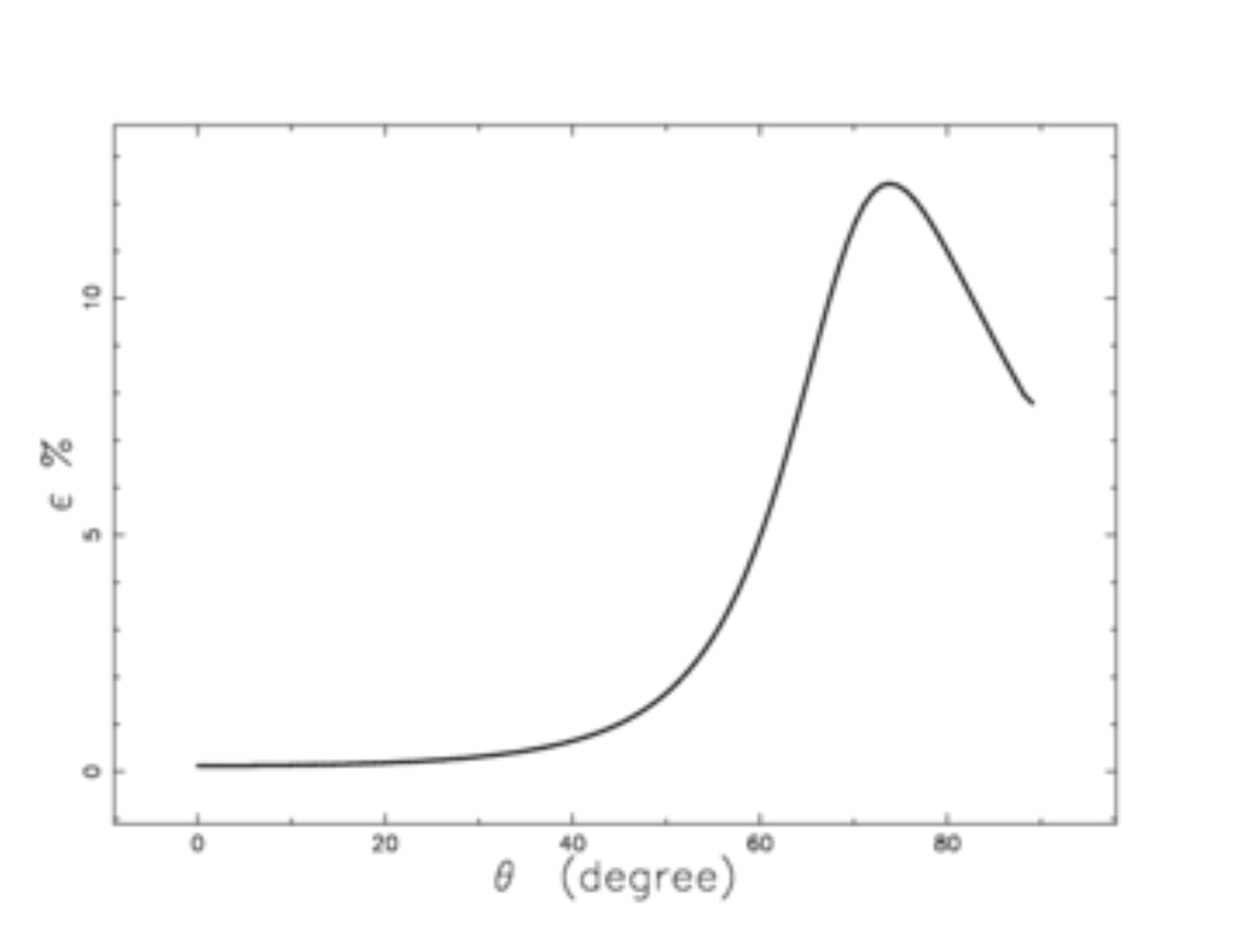}
\end {center}
\caption
{
Percentage error
of the Pad\'e  approximant solution compared
to the numerical solution, as a function of the
angle $\theta$. 
Other  parameters as in Figure \ref{pade_exp}, 
exponential profile.
}
\label{pade_effi_exp}
    \end{figure*}

\begin{figure*}
\begin{center}
\includegraphics[width=7cm]{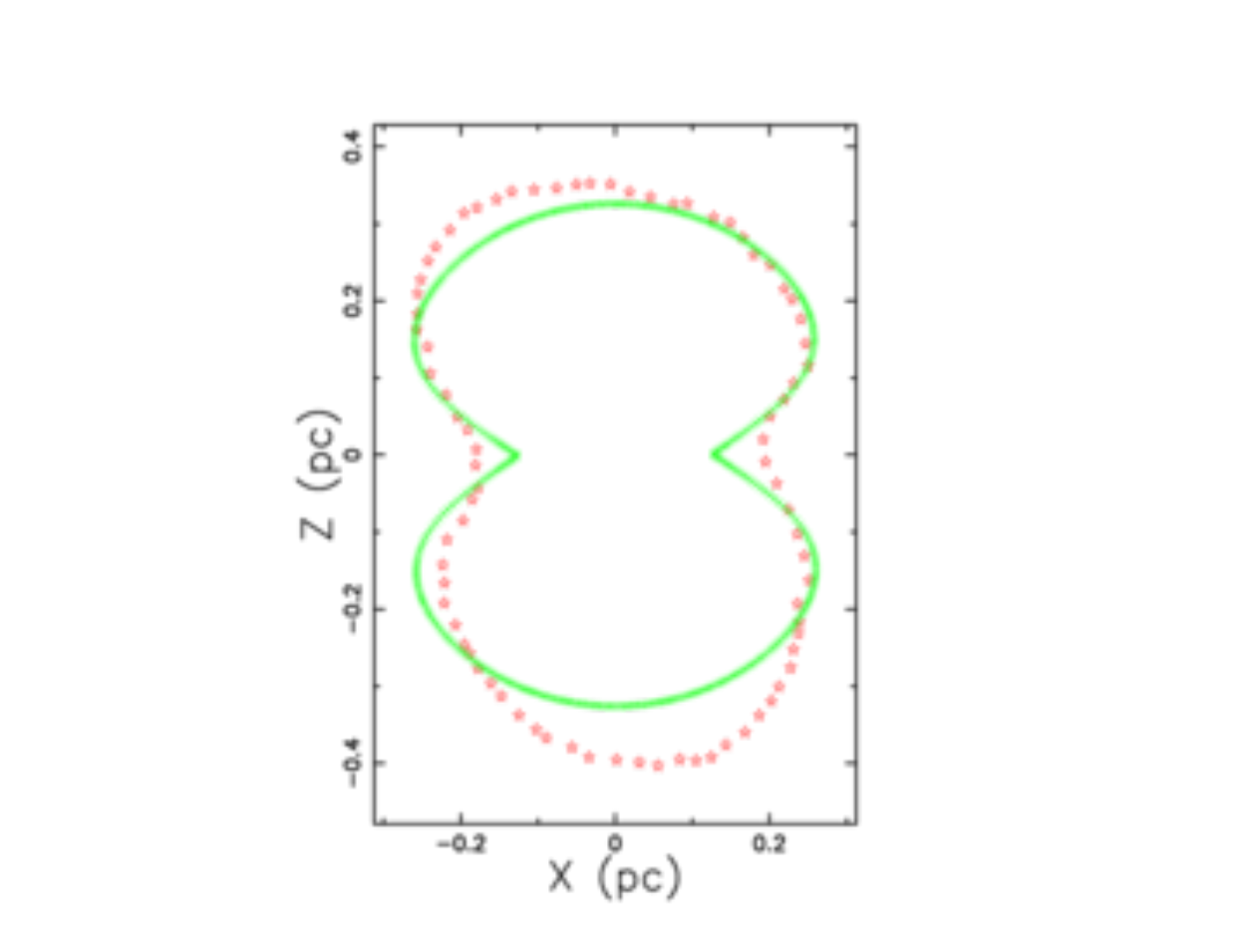}
\end {center}
\caption
{
Geometrical section of \sn1987a
in the $x-z$ plane with an exponential profile
(green points)
and observed profile
(red stars).
The parameters are the same  as Figure \ref{pade_exp}
and
$\epsilon_{\mathrm {obs}}=90.66\%$.
}
\label{cut_exp_1987a}
    \end{figure*}

In the case  of an {\it exponential}
density profile,
see  equation (\ref{profexponential}),
the differential equation which regulates the
motion is equation~(\ref{eqndiffexp}).
Figure \ref{pade_exp} compares
the numerical solution,
the approximate series solution,
and the  Pad\'e  approximant solution.
The above figure clearly shows the limited range of validity
of the power series  solution.
The good agreement between  the Pad\'e  approximant solution
and numerical solution, in
Figure \ref{pade_exp}
the two solutions  can  not
be distinguished,
has a percentage error
\begin{equation}
\epsilon = \frac{\big | r(t) - r(t)_{2,1} \big |}
{r(t) } \times 100
\quad ,
\end{equation}
where $r(t)      $ is the numerical solution
and   $r(t)_{2,1}$ is the Pad\'e  approximant solution.
Figure \ref{pade_effi_exp} shows
the percentage error as a function of the
polar angle $\theta$.
Figure \ref{cut_exp_1987a} shows a cut  of  \sn1987a
in the $x-z$ plane evaluated with the numerical solution.

\begin{figure*}
\begin{center}
\includegraphics[width=7cm]{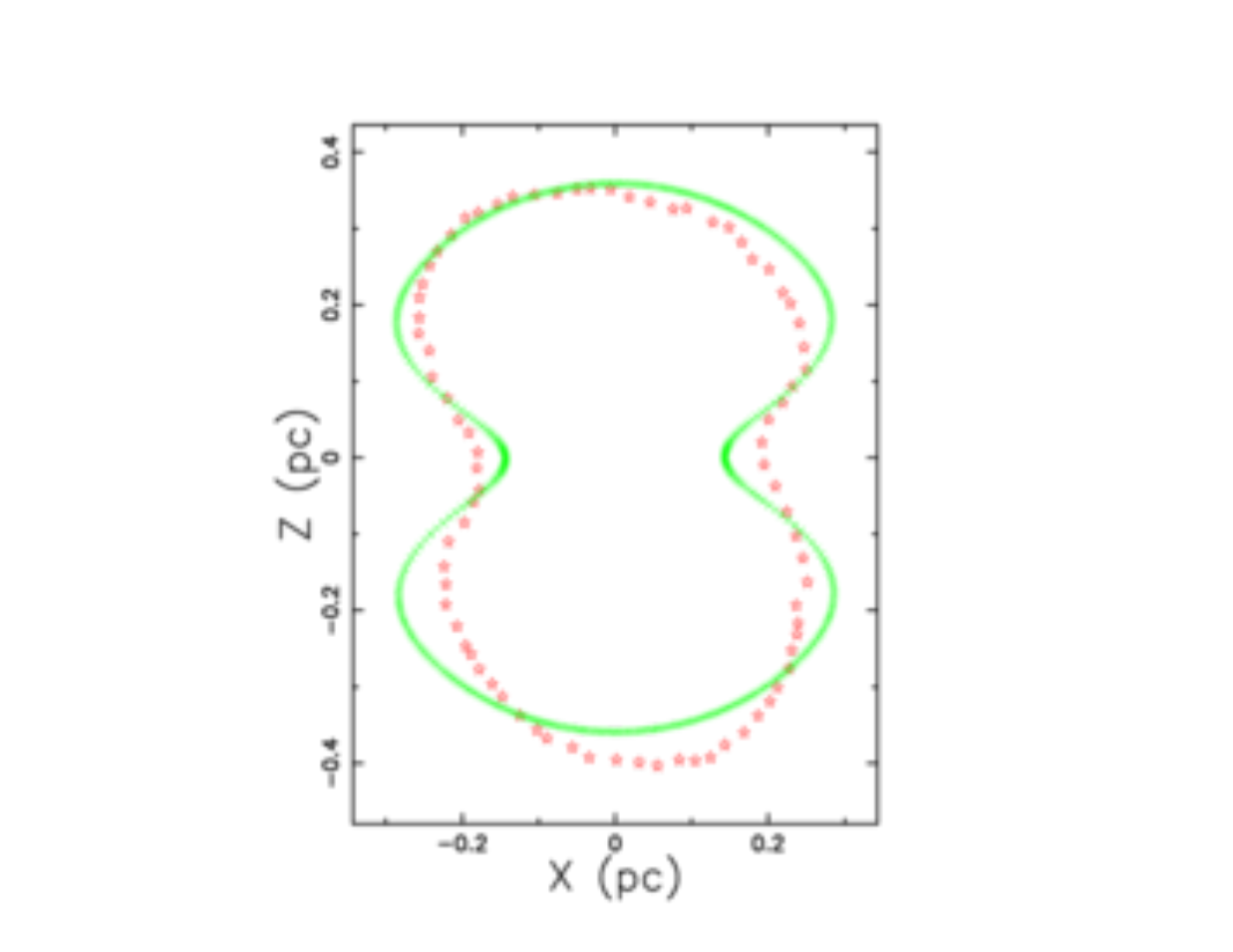}
\end {center}
\caption
{
Geometrical section of \sn1987a
in the $x-z$ plane with a Gaussian profile
(green points)
and observed profile
(red stars).
The parameters
$r_0=0.07$ pc,
$t=21.86$\ yr,
$t_0 =0.1$\ yr,
$b= 0.018$ pc,
$v_0=13000$\ km s$^{-1}$
give
$\epsilon_{\mathrm {obs}}=90.95\%$.
}
\label{cut_gauss_1987a}
    \end{figure*}

\begin{figure*}
\begin{center}
\includegraphics[width=7cm]{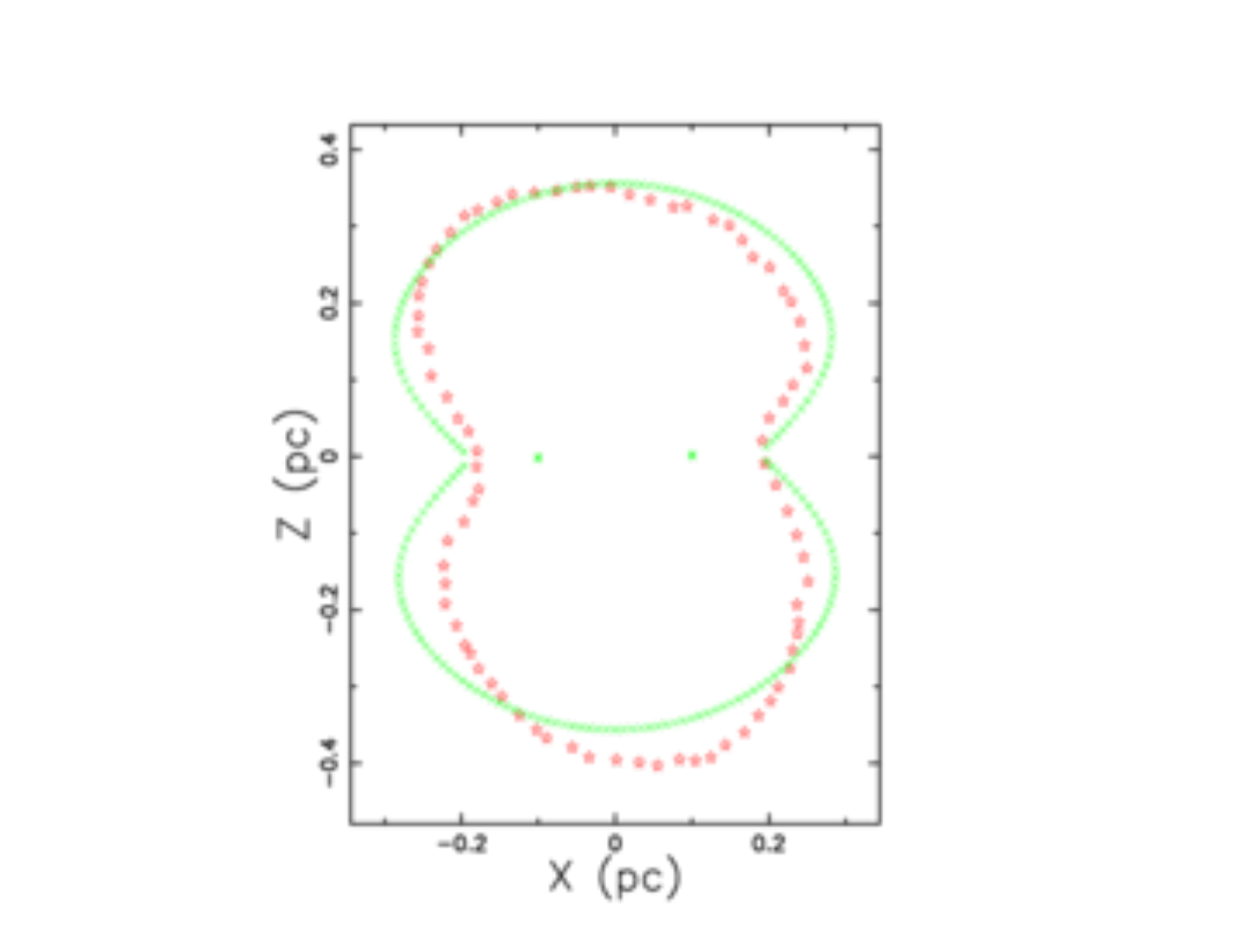}
\end {center}
\caption
{
Geometrical section of \sn1987a
in the $x-z$ plane with an exponential
profile: relativistic case
(green points)
and observed profile (red stars).
The parameters
$r_0$=0.1  pc,
$b=0.02$\ pc,
$t=21.86$\ yr,
$t_0 =0.1$~yr,
$v_0\,=70000$\ km s$^{-1}$,
and $\beta_0=0.233$
give
$\epsilon_{\mathrm {obs}}=90.23\%$.
}
\label{cut_exp_rel_1987a}
    \end{figure*}

In the case  of a  {\it Gaussian}
density profile,
see  equation (\ref{profgaussian}),
the differential equation~(\ref{eqndiffgauss}).
regulates the
motion.
Figure \ref{cut_gauss_1987a}
shows a cut  of  \sn1987a
in the $x-z$ plane evaluated with a numerical solution
for the Gaussian profile.
In this case the scale of the 
Gaussian profile is 0.018
and therefore 4194 time smaller 
of  scale of  
the galactic H\,I  ,
see equation (\ref{galactich}).
In the case  of an {\it exponential}
density profile,
see  equation (\ref{profexponential}),
the differential 
equation (\ref{eqndiffrelexp})  regulates the
relativistic motion.
Figure \ref{cut_exp_rel_1987a} shows the numerical solution
for  \sn1987a
in the $x-z$ plane for the {\it relativistic-exponential} case.

\subsection{The case \s1006}

\begin{figure*}
\begin{center}
\includegraphics[width=7cm]{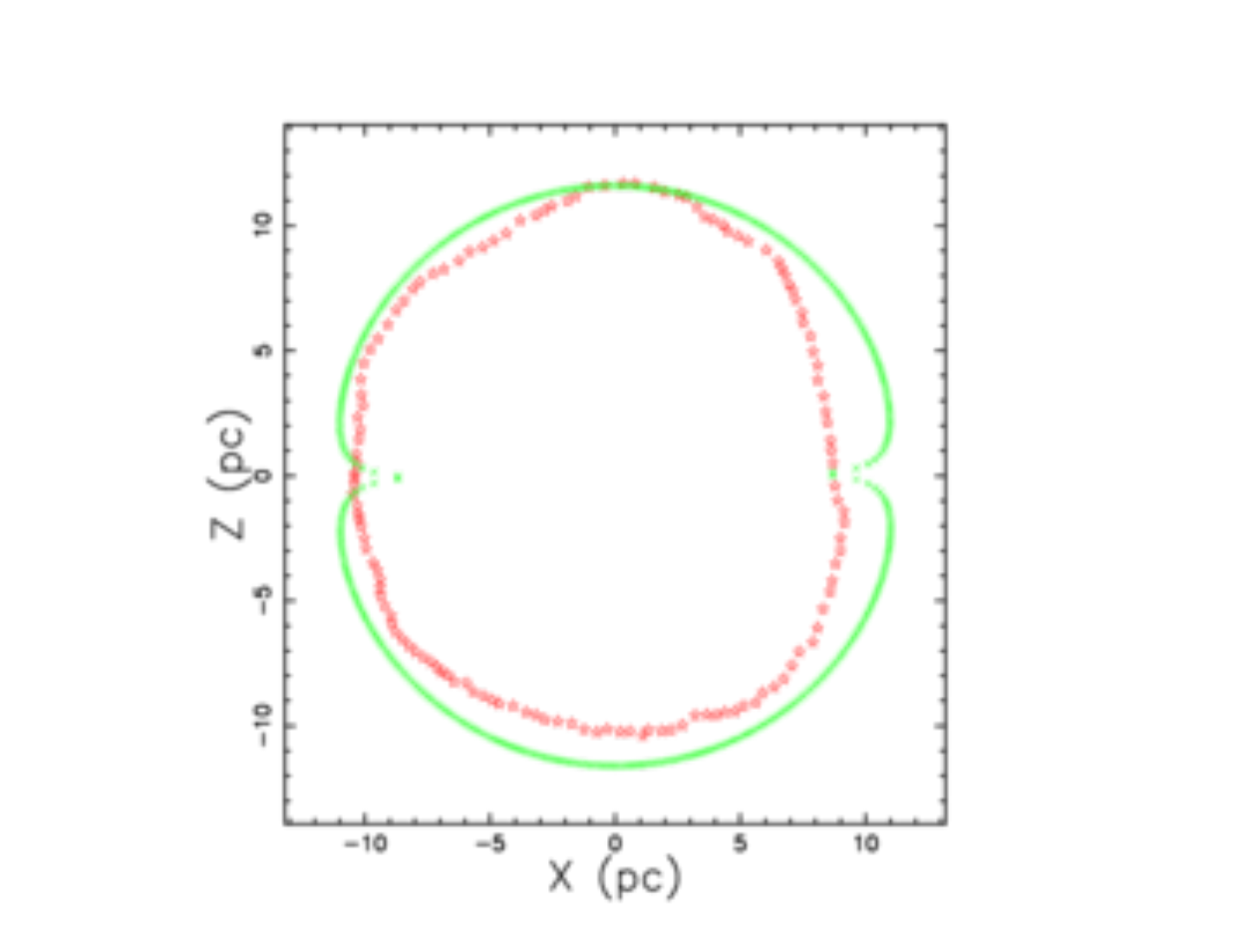}
\end {center}
\caption
{
Geometrical section of \s1006
in the $x-z$ plane with an hyperbolic profile
(green points)
and observed profile (red stars).
The parameters
$r_0=1$\ pc,
$z_0=0.00015$\ pc,
$t=1000$\ yr,
$t_0 =0.1$ yr  and
$v_0\,=10600$ km s$^{-1}$
give
$\epsilon_{\mathrm {obs}}=89.55\%$.
}
\label{sn1006_vera_obs_hyper}
    \end{figure*}

\begin{figure*}
\begin{center}
\includegraphics[width=7cm]{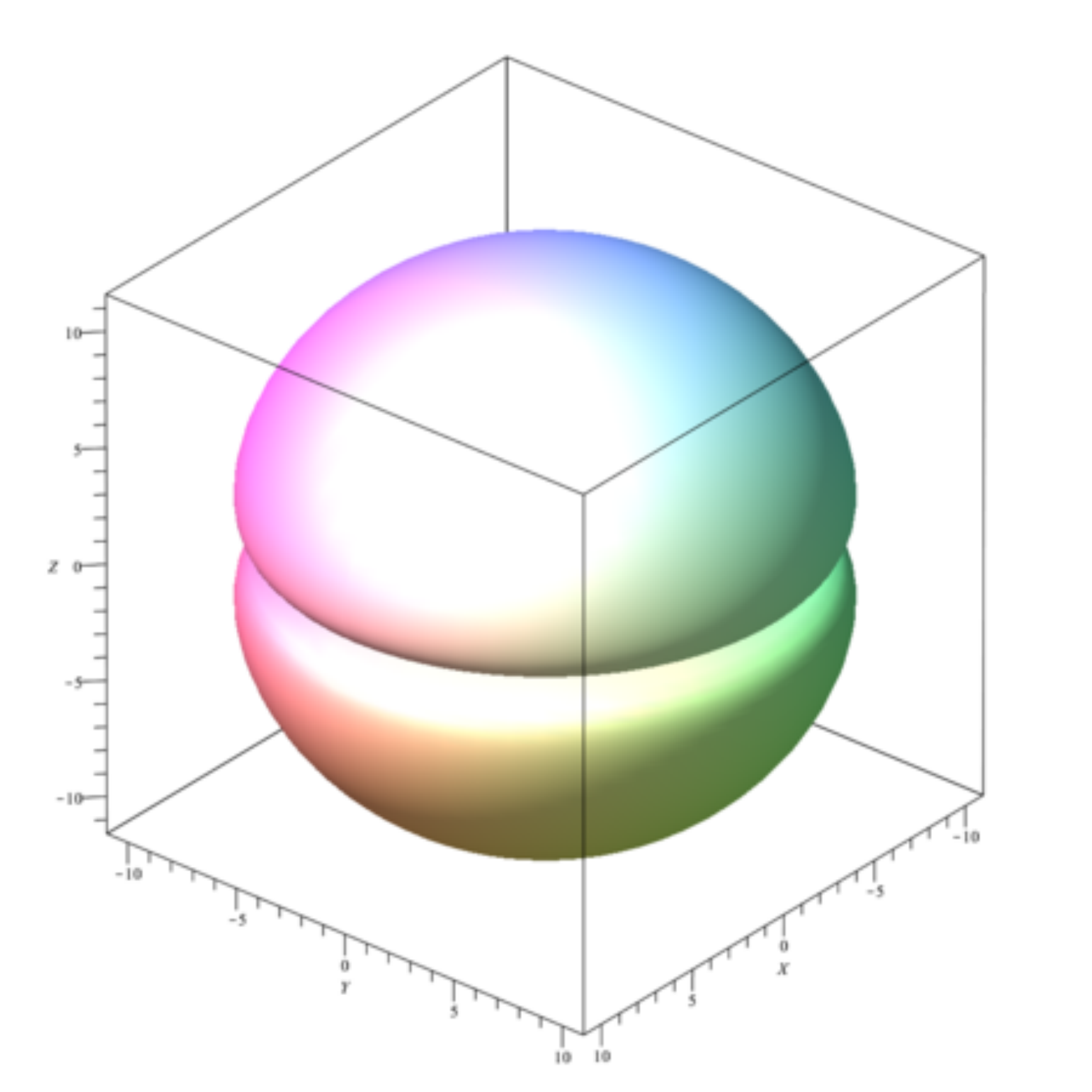}
\end {center}
\caption
{
3D surface  of  \s1006
with parameters as in Figure \ref{sn1006_vera_obs_hyper}, 
hyperbolic profile.
The three Euler angles are $\Theta=40$, $\Phi=60$ and
$ \Psi=60 $.
}
\label{3dsurfacehypersn1006}
    \end{figure*}

In the case  of an {\it hyperbolic}
 density profile,
Figure \ref{sn1006_vera_obs_hyper} displays a cut  of  \s1006
and Figure \ref{3dsurfacehypersn1006}
a 3D surface.

In presence of an exponential density profile
for the CSM
as given by equation (\ref{profexponential}),
in the NCD case we solve numerically 
the differential equation (\ref{eqndiffexpncd})
with parameters as in Figure (\ref{datafit1006}).
\begin{table}
\caption
{
Numerical value of the parameters
of the simulation  for  \s1006, exponential profile and NCD case.
}
\label{datafit1006}
 \[
 \begin{array}{lc}
 \hline
 \hline
 r_0              &   0.09 \,pc         \\
 b                &   3\times r_0       \\
 v_0              &   8500 \frac{km}{s} \\
p                 &   30    \\
t_{0,1}           &   10   \,yr \\
t_{1}             &   1000 \,yr  \\
\noalign{\smallskip}
 \hline
 \hline
 \end{array}
 \]
 \end {table}

\begin{figure}
\includegraphics[width=6cm]{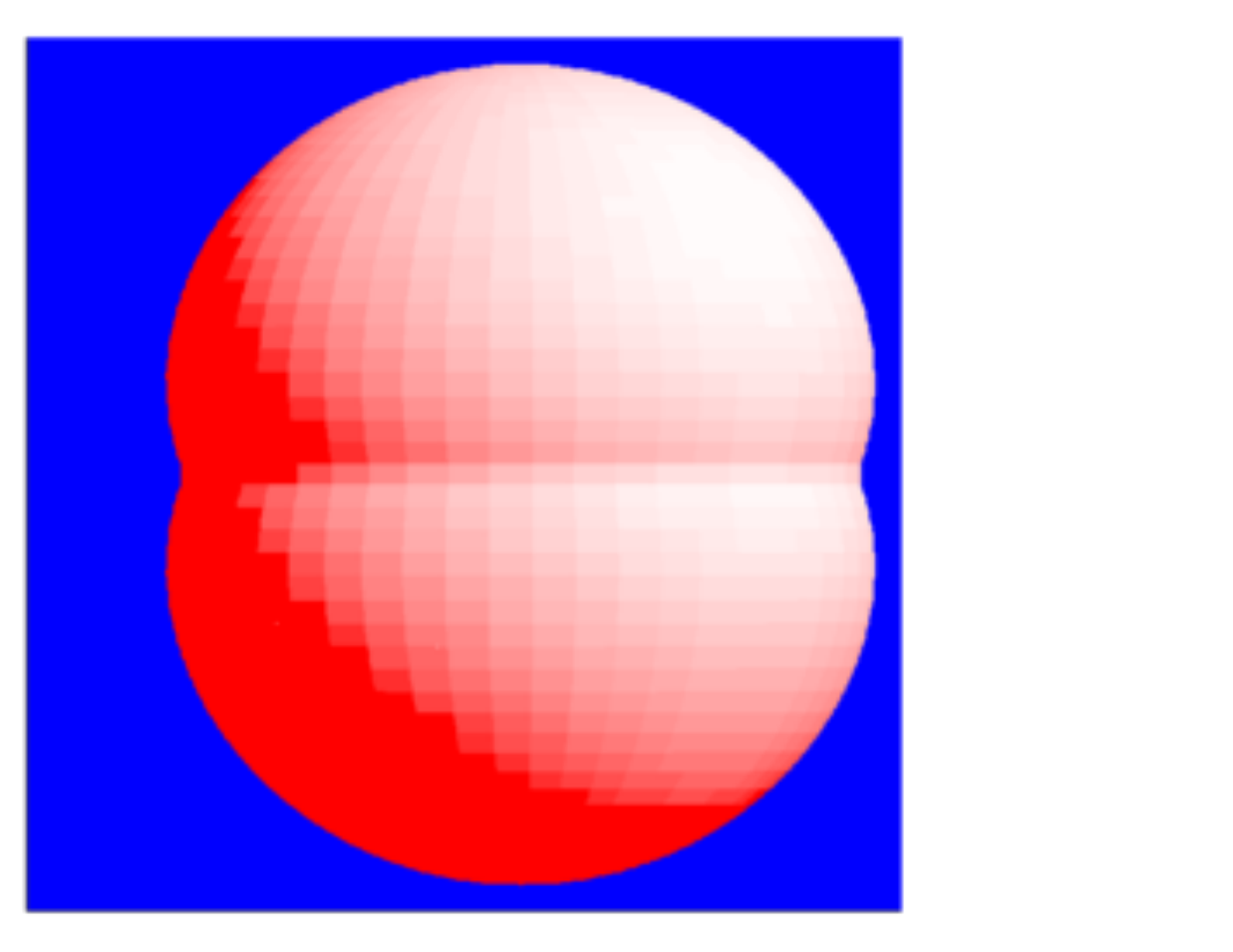}
\caption {
Continuous  three-dimensional surface of 
\s1006 : the three Euler angles
characterizing the point of
view are
     $ \Phi   $=90    $^{\circ }  $,
     $ \Theta $=90    $^{\circ }  $
and  $ \Psi   $=90    $^{\circ }  $.
Physical parameters as in Table \ref{datafit1006}, 
exponential profile and NCD case.
          }%
    \label{1006_faces}
    \end{figure}

\begin{figure*}
\begin{center}
\includegraphics[width=7cm]{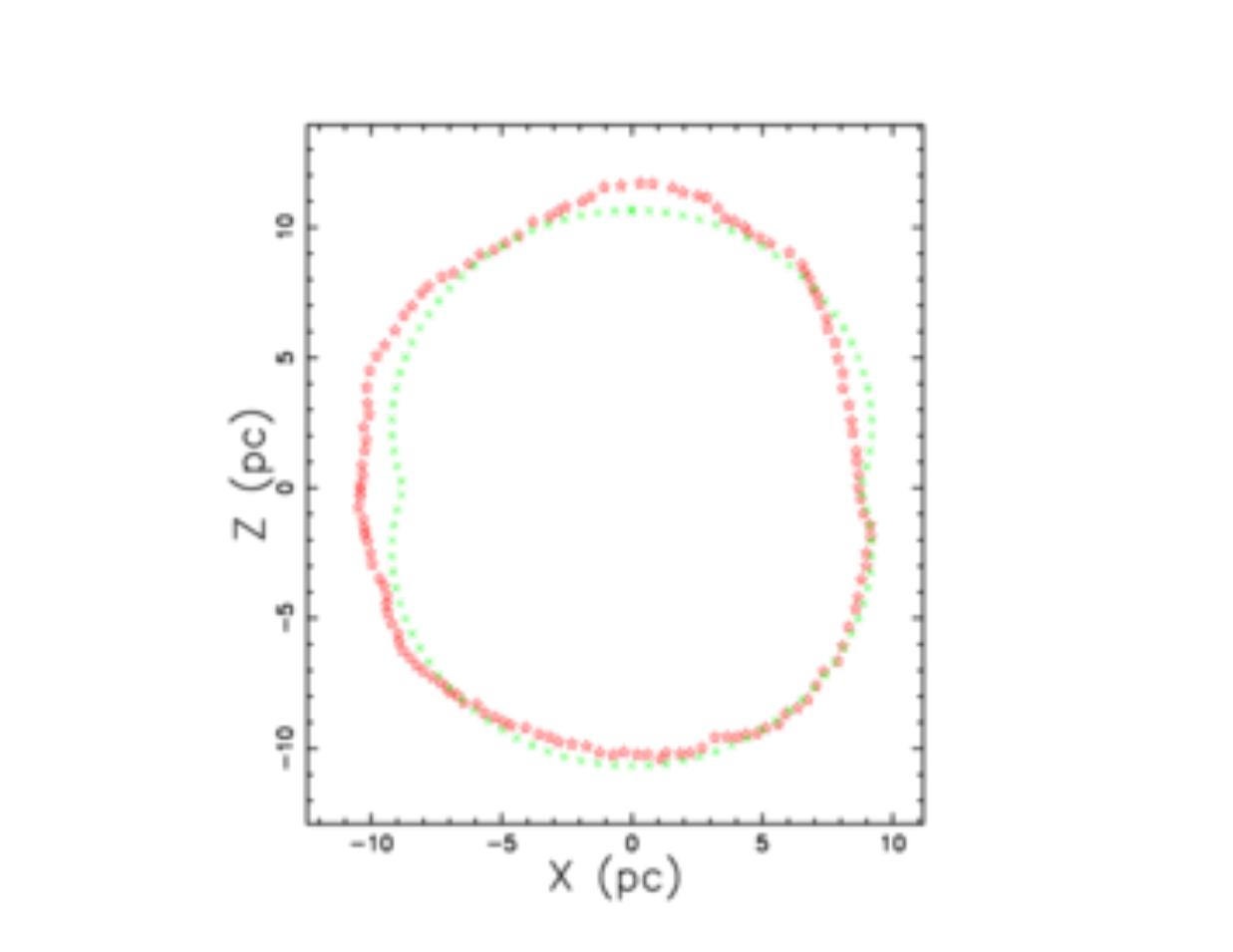}
\end {center}
\caption
{
Geometrical section of \s1006
in the $x-z$ plane with an exponential profile
(green points)
and observed profile
(red stars).
Parameters as  in Table \ref{datafit1006}
and efficiency along the equatorial direction
$\epsilon_{\mathrm {eq}}=98.25\%$.
}
\label{sn1006_vera_obs}
    \end{figure*}

The weakly asymmetric 3D shape  of  \s1006
is reported  in Figure \ref{1006_faces}.
Our model for  \s1006 predicts a minimum velocity  in the
equatorial plane  of 5785~km/s  and  a maximum velocity of 7229~km/s  in the polar direction. A recent observation of \s1006
quotes a minimum velocity of
 5500~km/s  and a maximum velocity of 14500~km/s   assuming a distance of 3.4 kpc.
Figure \ref{sn1006_vera_obs} shows a cut  of  \sn1987a
in the $x-z$ plane evaluated with the numerical solution.

\section{Symmetrical synchrotron radiation}

This section introduces
the transfer equation,  
the acceleration and losses for electrons,
the mathematical diffusion in 3D,
a simple model for the image formation,
and a diffusive model.

\subsection{The transfer equation}

The transfer equation in the presence of emission only,
see, for example,  
\cite{rybicki}
 or
\cite{Hjellming1988},
 is
 \begin{equation}
\frac {dI_{\nu}}{ds} =  -k_{\nu} \rho I_{\nu}  + j_{\nu} \rho
\label{equazionetrasfer}
\quad ,
\end {equation}
where  $I_{\nu}$ 
is the specific intensity, 
$s$ is the
line of sight, 
$j_{\nu}$ is the emission coefficient,
$k_{\nu}$ is a mass absorption coefficient,
$\rho$ is the density of mass at position $s$,
and the index $\nu$ denotes the frequency of
emission.
The intensity of radiation, i.e. the photon flux,
is here identified with the counts at a given energy.
The solution to Equation~(\ref{equazionetrasfer})
 is
\begin{equation}
 I_{\nu} (\tau_{\nu}) =
\frac {j_{\nu}}{k_{\nu}} ( 1 - e ^{-\tau_{\nu}(s)} )
\quad  ,
\label{eqn_transfer}
\end {equation}
where $\tau_{\nu}$ is the 
optical depth at frequency $\nu$:
\begin{equation}
d \tau_{\nu} = k_{\nu} \rho ds
\quad.
\end {equation}
The framework of synchrotron emission,
as described in  sec. 4 of \cite{Schlickeiser},
 is 
often used in order to explain an SNR,
see for example 
\cite{Yamazaki2014,Tran2015,Katsuda2015,Dubner2015,Reynolds2017}.
The volume emissivity 
(power 
per unit frequency interval 
per unit volume per unit solid angle) 
of the ultrarelativistic  electrons, 
according to 
\cite{lang},  
 is  
\begin {equation}
\epsilon (\nu) =
\int  P(\nu) N(E) dE
\quad ,
\end{equation}
where $P(\nu)$ is the total power radiated 
per unit frequency interval by one electron
and  $N(E) dE$ is the number of electrons 
per unit  volume,  
per unit solid angle along the line
of sight, which are moving in the direction 
of the observer and whose energies lie in the range
$E$ to $E+dE$.
In the case of a power law  spectrum,
\begin{equation}
N(E)dE = K E^{-\gamma} dE  
\label{spectrum}
\quad  ,
\end{equation}
where $K$ is a constant.  
The value  of the constant  $K$ can be found 
by assuming that  
the  probability density function 
(PDF, in the following)
for relativistic energy  
is of  Pareto type   
as defined in   \cite{evans}:
\begin {equation}
f(x;a,c) = {c a^c}{x^{-(c+1)}} \quad ,
\label{pareto}
\end {equation}
with $ c~> 0$.
In our case, 
$c=\gamma-1$ and  $a=E_{min}$,
where $E_{min}$ is the minimum energy.
In the previous formula, we can extract
\begin{equation}
K= 
N_0 (\gamma -1) E_{min}^{\gamma -1} 
\quad  ,
\end{equation}
where $N_0$ is the total number
of relativistic electrons  per unit volume, 
here assumed to be approximately 
equal to the matter number density.
The previous  formula can also be expressed 
as  
\begin{equation}
K= 
\frac {\rho}{1.4 \, m_{\mathrm {H}}}
 (\gamma -1) E_{min}^{\gamma -1} 
\quad  ,
\end{equation}
where $m_H$ is the mass of hydrogen.
The emissivity  of  the ultrarelativistic  
electrons  from a homogeneous  
and isotropic distribution of electrons 
whose  $N(E)$  is given by 
Equation~(\ref{spectrum})
is, according to 
\cite{lang},  
\begin{eqnarray}
j_{\nu} \rho  =  \\
\approx 0.933 \times 10^{-23}
\alpha_L (\gamma) K H_{\perp} ^{(\gamma +1)/2 }
\bigl (
 \frac{6.26 \times 10^{18} }{\nu}
\bigr )^{(\gamma -1)/2 } \nonumber    \\
\mathrm{erg sec}^{-1} \mathrm{cm}^{-3} \mathrm{Hz}^{-1} \mathrm{rad}^{-2}
\nonumber   
\quad , 
\end{eqnarray}
where $\nu$ is the frequency
and   $\alpha_L (\gamma)$  is a slowly
varying function 
of $\gamma$ which is of the order of unity 
and is given by
\begin{equation}
\alpha_L(\gamma) =
2^{(\gamma -3)/2} \frac{\gamma+7/3}{\gamma +1}
\Gamma \bigl ( \frac {3\gamma -1 }{12} \bigr )  
\Gamma \bigl ( \frac {3\gamma +7 }{12} \bigr )  
\quad ,
\end{equation} 
for  $\gamma \ge \frac{1}{2}$. 

We now continue to analyse a first  case 
of an optically thin layer
in which $\tau_{\nu}$ is very small
(or $k_{\nu}$ is very small)
and where the density $\rho$ is replaced
by  the concentration $C(s)$
 of relativistic electrons:
\begin{equation}
j_{\nu} \rho =K_e  C(s)
\quad  ,
\end{equation}
where $K_e$ is a constant function
of the energy power law index,
magnetic field,
and frequency of e.m.  emission.
The intensity is now
\begin{equation}
 I_{\nu} (s) = K_e
\int_{s_0}^s   C (s\prime) ds\prime \quad  \mbox {optically thin layer}.
\label{transport}
\end {equation}
The increase in brightness
is proportional to the concentration 
integrated along
the line of  sight.
A second case  analyses a  
quadratic  relationship between  
the  emission coefficient and  the concentration
\begin{equation}
j_{\nu} \zeta =K_2  C(s)^2
\quad  ,
\label{eqn_transfer_square}
\end{equation}
where $K_2$ is a  constant function.
This is true for 
\begin{itemize} 
\item 
Free-free radiation from a thermal plasma,
see formula (1.219) in  \cite{lang}  .
\item 
Thermal bremsstrahlung and recombination radiation ,
see formula (1.237) in  \cite{lang}  .
\end{itemize}
The  intensity in the "thermal case" is 
\begin{equation}
 I_{\nu} (s) = K_2 
\int_{s_0}^s   C (s\prime)^2 ds\prime \quad  \mbox {optically thin layer}
\quad quadratic~case \quad .
\label{transport2}
\end {equation}

\subsection{Acceleration and losses for  electrons}

An electron which  loses
energy  due to 
synchrotron radiation
has a lifetime  of
\begin{equation}
\tau_r  \approx  \frac{E}{P_r} \approx  500  E^{-1} H^{-2} sec
\quad ,
\label {taur}
\end{equation}
where
$E$  is the energy in ergs,
$H$ the magnetic field in gauss,
and
$P_r$  is the total radiated
power, see Eq. 1.157 in  \cite{lang}.
The  energy  is connected  to  the critical
frequency, see Eq. 1.154 in  \cite{lang},
by
\begin {equation}
\nu_c = 6.266 \times 10^{18} H E^2~\mathrm{Hz}
\quad  .
\label {nucritical}
\end{equation}
\index{critical frequency}
The lifetime
for synchrotron  losses is
\begin{equation}
\tau_{syn} =
 39660\,{\frac {1}{H\sqrt {H\nu}}} \, \mathrm{yr}
\quad  .
\end{equation}
\index{synchrotron  losses}
A first  astrophysical  result  is the evaluation of  the distance,
$L_{syn}$,
that a ultrarelativistic  electron which radiates at a given 
frequency  $\nu$ when is moving 
with the advancing shell  at  a given  average velocity, $v_{SN}$,
\begin{equation}
L_{syn} =  12142.95144\,{\frac {{\it \beta_{SN}}}{H\sqrt {H\nu}}}
\quad  pc 
\quad  ,
\end{equation}
where  $\beta_{SN} = \frac{v_{SN}}{c}$.
As a  practical  example  $L_{syn}=3.84\,10^{-3}$~pc
when $\beta_{SN}=1/100$, $\nu=1.0\,10^{9}$~Hz and 
$H=1$~Gauss.

Considering a given radius $R_{SN}$ of the advancing SN,
we should  speak of the `in situ' acceleration of the electrons
when 
\begin{equation}
R_{SN} \,> L_{syn}
\quad .
\label{insitu}
\end{equation}   
In the framework of the `in situ' acceleration of the electrons 
and following \cite{Fermi49,Fermi54},
the gain  in  energy  in a continuous    form
for  a particle
which  spirals  around a line of force
is  proportional to its
energy, $E$,
\begin  {equation}
\frac {d  E}  {dt }
=
\frac {E }  {\tau_{II} }   \quad,
\end {equation}\index{particle acceleration}
where $\tau_{II}$ is the  typical time scale,
\begin {equation}
\frac{1}{\tau_{II}}  = \frac {4} {3 }
( \frac {u^2} {c^2 }) (\frac {c } {L_{II} })
\quad ,
\label {tau2}
\end   {equation}
where $u$ is the velocity of the accelerating cloud
belonging  the advancing shell of the SN/SNR,
    $c$    is the speed
of light,
and $L_{II}$  is the
mean free path between clouds, 
see Eq. 4.439 in \cite{lang}. \index{Fermi!I}
The mean free path between the accelerating clouds
in the Fermi II mechanism can be found from the following
inequality in time:
\begin{equation}
\tau_{II} < \tau_{sync}
\quad  ,
\end{equation}
which  corresponds to  the following  inequality for the
mean free  path  between scatterers
\begin{equation}
L  <
\frac
{
1.72\,10^5 \,{u}^{2}
}
{
H\sqrt {H\nu}{c}^{2}
}
\,\mathrm{pc}
\quad  .
\end{equation}
The mean free path length for an SN/SNR
which emits synchrotron emission in the radio region   
at   $1.0\,10^{9}$ Hz,  
gives
\begin{equation}
L <
\frac
{
5.4618\,10^{-4}\,{{  u_{3000}}}^{2}
}
{
{{   H_1}}^{3/2}
}
\, \mathrm{pc}
\end{equation}
where  
$u_{3000}$ is the   velocity of the accelerating  
cloud  
expressed in units of 
3000 km/s
and  $H_1$  is the  magnetic field expressed
in   units of 1  gauss.
When this inequality  is fulfilled, the direct
conversion of the rate   of  kinetic energy
into radiation can be adopted.
Recall that the Fermi II  mechanism  produces
an  inverse power law
spectrum in the
energy
of the type
$
N (E) \propto  E ^{-\gamma}
$
or an  inverse power law  in the observed frequencies
$
N (\nu) \propto   \nu^{\beta }
$
with  $ \beta= - \frac{\gamma -1}{2}$
\cite {lang,Zaninetti2011a}.

The strong shock accelerating mechanism  
referred to as  Fermi I
was  introduced by  
\cite{Bell_I}
and 
\cite{Bell_II}.
The energy  gain  relative to a particle that is crossing
the shock is
\begin  {equation}
\frac {d  E}  {dt }
=
\frac {E }  {\tau_{I} }   \quad,
\end {equation}
where $\tau_{I}$ is the  typical time scale,
\begin {equation}
\frac{1}{\tau_{I}}  = \frac {2} {3 }
( \frac {u} {c }) (\frac {c } {L_{I} })
\quad ,
\end   {equation}
where $u$ is the velocity of the shock of the SN/SNR,
$c$  is the speed
of light,
and $L_{I}$  is the
mean free path between scatterers.\index{Fermi!II}
This process produces an energy spectrum of
electrons
of the type
\begin {equation}
N (E) dE \propto E^{-2} dE
\quad,
\end   {equation}
see  \cite{longair}.
The  two mechanisms,
Fermi I and Fermi II, 
produce the  same results  when
\begin{equation}
\frac{L_{II}} {L_{I}} =
2 \frac{u}{c}
\quad .
\end{equation}
More details can be found in \cite{Zaninetti2012g}.

\subsection{3D diffusion from a spherical source}

\label{mathematical}
Once the number density, $C$, and the diffusion coefficient, $D$,
are introduced,  Fick's first equation
changes   its expression on the basis  of the adopted
environment,  see for example equation~(2.5) in  \cite{berg}.
In three dimensions, it is
\begin{equation}
\frac {\partial C }{\partial t} =
D \nabla^2 C
\quad,
\label{eqfick}
\end {equation}
where $t$ is the time  and $ \nabla^2$ is
the Laplacian differential operator.
In the presence of the  steady state condition,
\begin{equation}
D \nabla^2 C   = 0
\quad .
\label{eqfick_steady}
\end {equation}

\begin{figure}[pb]
\includegraphics[width=7cm]{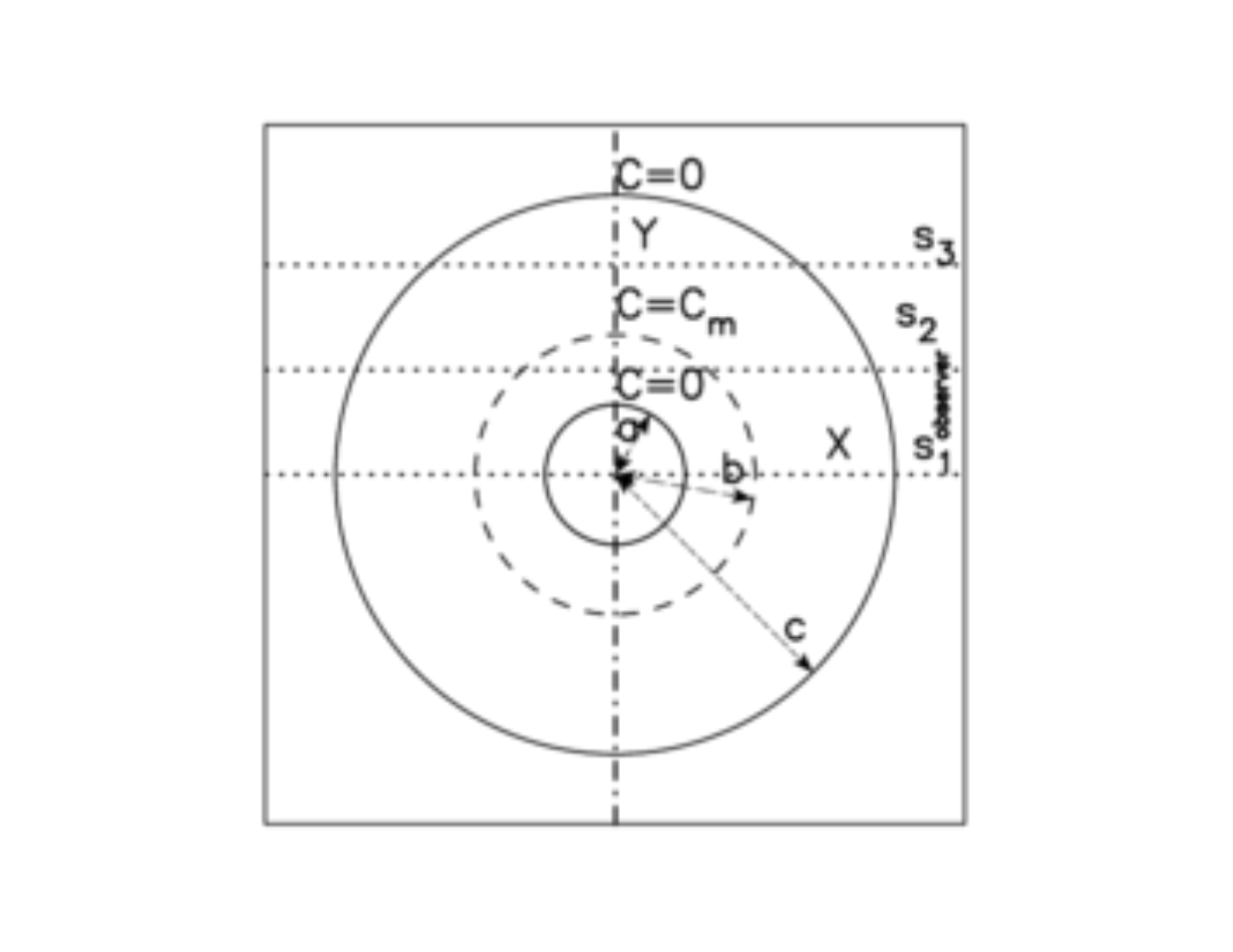}
\vspace*{8pt}
\caption {
The spherical source is  represented  by 
a dashed line, the two absorbing boundaries
with a full line.
\label{plotdiffabc}}
 \end{figure}

The  number density rises from 0 at {   r=a}  to a
maximum value $C_m$ at {   r=b} and then  falls again
to 0 at {    r=c}, 
see Figure \ref{plotdiffabc}, 
which is adapted from Fig. 3.1 of~\cite{berg}.

The  solution to  equation~(\ref{eqfick_steady})
 is
\begin{equation}
C(r) = A +\frac {B}{r}
\quad,
\label{solution}
\end {equation}
where $A$ and $B$  are determined by  the boundary conditions,
\begin{equation}
C_{ab}(r) =
C_{{m}} \left( 1-{\frac {a}{r}} \right)  \left( 1-{\frac {a}{b}}
 \right) ^{-1}
\quad a \leq r \leq b
\quad,
\label{cab}
\end{equation}
and
\begin{equation}
C_{bc}(r)=
C_{{m}} \left( {\frac {c}{r}}-1 \right)  \left( {\frac {c}{b}}-1
 \right) ^{-1}
\quad b \leq r \leq c
\quad.
\label{cbc}
\end{equation}
These solutions can be found in
\cite{berg}
or in
\cite{crank}.

\subsection{Spherical  Image}

\begin{figure*}
\includegraphics[width=6cm]{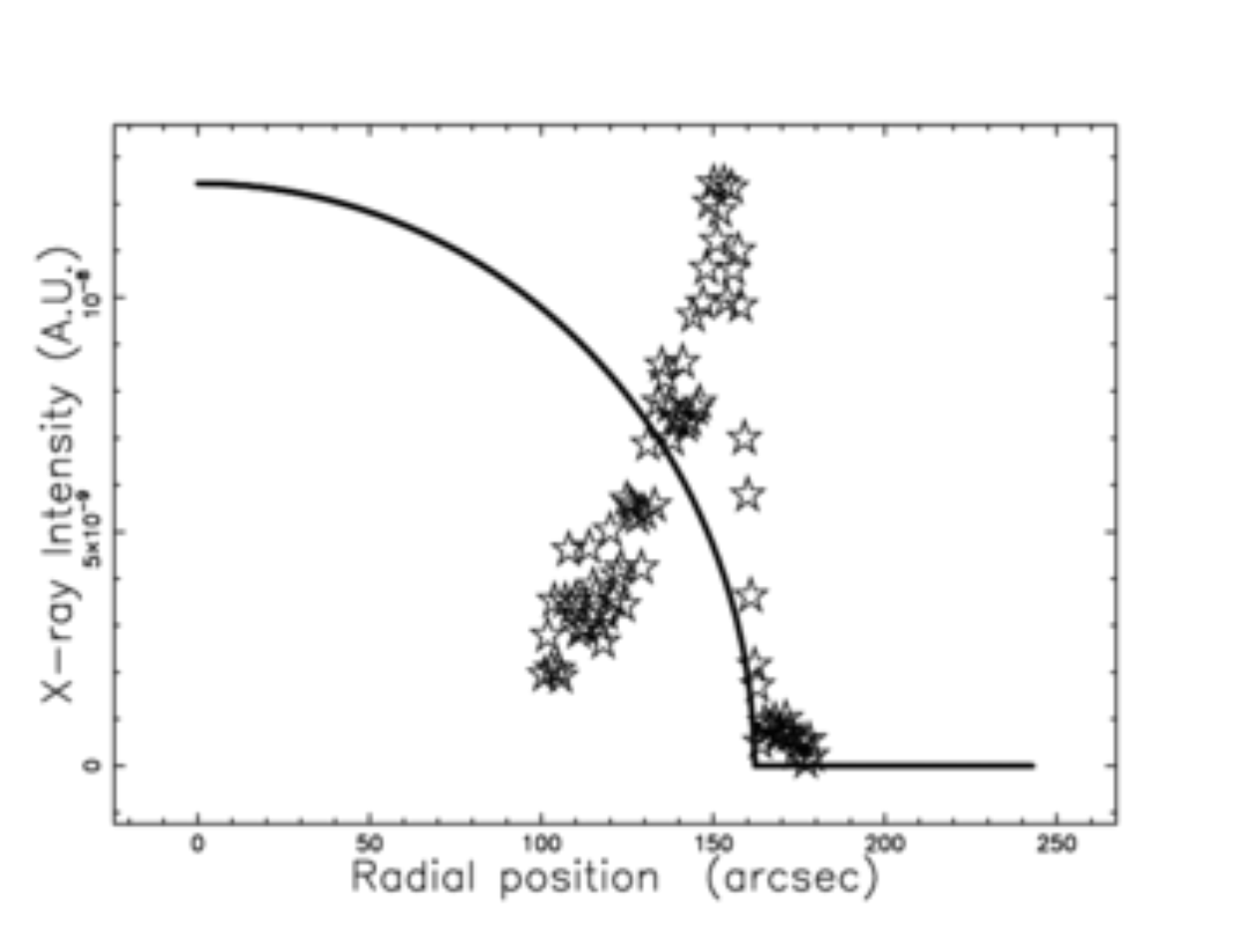}
\caption {
 Cut of the thermal   intensity ${   I}$
 of the rim  model (equation~(\ref{isphere}))
 through the centre  (full  line) of \s1006
 and  real X-data  (empty stars)
 when  $b=162\,arcsec$.
 The observed data  been extracted  by the author
 from Fig.~8 (subfigure 14) of
 \cite{Ressler2014}.
} \label{sn1006_thermal}
    \end{figure*}

A {\it first} thermal model for the image   is characterized by a
constant temperature in  the internal region of the advancing
sphere. 
We recall  that in the thermal model the temperature 
and the density are constant  in the advancing shell,
see equation (10.29) in \cite{McCray1987}.
We therefore assume that the number density $C$ is
constant in a sphere of radius $a$ and then falls  to 0. The
line of sight, when the observer is situated at the infinity of
the $x$-axis, is the locus parallel to the $x$-axis which
crosses  the position $y$ in a Cartesian $x-y$ plane and
terminates at the external circle of radius $a$, see
\cite{Zaninetti2009a}. The length  of this locus  is
\begin{eqnarray}
l_{ab} = 2 \times ( \sqrt {a^2 -y^2}) \quad  ;   0 \leq y < a
\quad . \label{lengthsphere}
\end{eqnarray}
The number density $C_m$ is constant  in the sphere of radius $a$
and therefore the intensity of the radiation is
\begin{eqnarray}
I_{0a} =C_m \times  2 \times ( \sqrt { a^2 -y^2})
 \quad  ;  0 \leq y < a    \quad .
\label{isphere}
\end{eqnarray}
A comparison   of the observed data of \s1006  with the theoretical
thermal intensity is displayed in Figure \ref{sn1006_thermal}.

\begin{figure*}
\includegraphics[width=6cm]{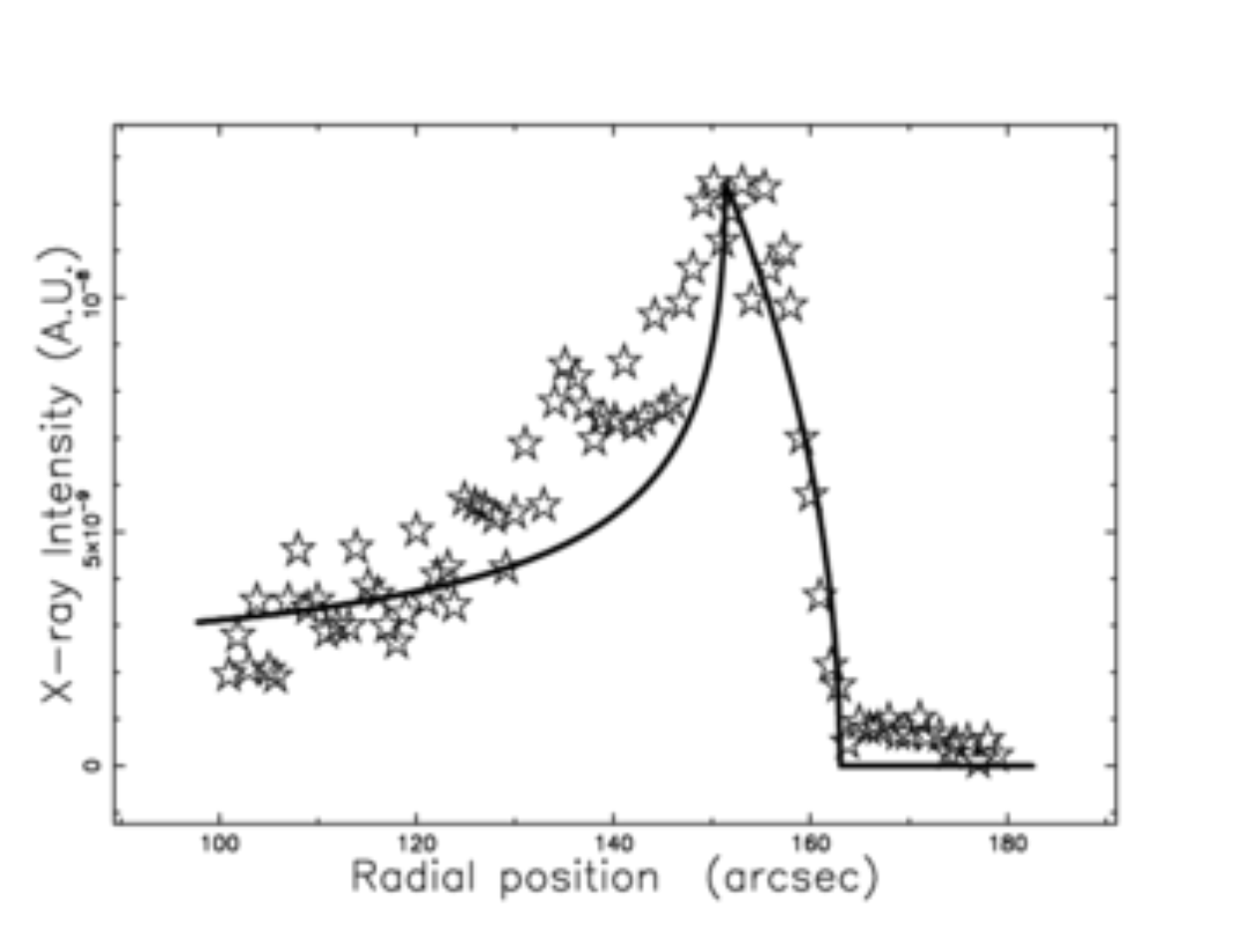}
\caption {
 Cut of the non-thermal   intensity ${   I}$
 of the rim  model (equation~(\ref{irim}))
 through the centre  (full  line) of \s1006 
 and  real X-data  (empty stars).
 The parameters  are
 $a=149.3 \,arcsec  $ pc and $b=160\,arcsec$.
 The observed data  has been extracted  by the author
 from Fig.~8 (subfigure 14) of
 \cite{Ressler2014}.
} \label{sn1006_const}
    \end{figure*}

A {\it second} non-thermal model for the image is characterized by
synchrotron emission 
by  ultrarelativistic electrons
in a thin layer around the advancing sphere. 
We therefore
assume that the number density $C$ is constant in the spherical
thin layer, and in particular
rises from 0 at $r=a$ to a maximum value $C_m$, remains constant
up to $r=b$ and then falls again to 0. The line of sight, when
the observer is situated at the infinity of the $x$-axis, is the
locus parallel to the $x$-axis which  crosses  the position $y$ in
a Cartesian $x-y$ plane and terminates at the external circle of
radius $b$, see \cite{Zaninetti2009a}.
 The length of this locus
is
\begin{eqnarray}
l_{0a} = 2 \times ( \sqrt { b^2 -y^2} - \sqrt {a^2 -y^2})
\quad  ;   0 \leq y < a  \nonumber  \\
l_{ab} = 2 \times ( \sqrt { b^2 -y^2})
 \quad  ;  a \leq y < b    \quad .
\label{length}
\end{eqnarray}
The number density $C_m$ is constant between two spheres of radii
$a$ and $b$ and therefore the intensity of radiation is
\begin{eqnarray}
I_{0a} =C_m \times 2 \times ( \sqrt { b^2 -y^2} - \sqrt {a^2
-y^2})
\quad  ;   0 \leq y < a  \nonumber  \\
I_{ab} =C_m \times  2 \times ( \sqrt { b^2 -y^2})
 \quad  ;  a \leq y < b    \quad .
\label{irim}
\end{eqnarray}
The ratio between the theoretical intensity at the maximum $(y=a)$
 and at the minimum ($y=0$)
is given by 
\begin{equation}
\frac {I_{0a} =(y=a)} {I_{0a} =(y=0)} = \frac {\sqrt {b^2 -a^2}} {b-a}
\quad .
\label{ratioteorrim}
\end{equation}
A comparison   of the observed data of \snr  and the theoretical non-thermal intensity is displayed  in Figure ~\ref{sn1006_const}.

The main result of this paragraph is that the intensity of the
thermal model which has  the maximum of the intensity at the
centre of an SNR does not  match the observed profiles. The
observed profiles in the radio intensity have the maximum value at the
rim, as predicted by the non-thermal model.

\subsection{Diffusive model}

The  concentration rises from 0 at {   r=a}  
to a maximum value
$C_m$ at {   r=b} and then  falls again to 0 at {    r=c}. 
The concentrations to be used are formulas 
(\ref{cab}) and (\ref{cbc})
once $r=\sqrt{x^2+y^2}$ is imposed; these two concentrations are
inserted in formula~(\ref{eqn_transfer}), which  represents the
transfer equation.
 The geometry of the phenomenon fixes
 three different zones ($0-a,a-b,b-c$) for the variable $y$,
 see~\cite{Zaninetti2007_c,Zaninetti2009a};
 the first segment, $I^I(y)$, is
\begin{eqnarray}
I^I(y)=
\nonumber  \\
2 {\frac {b{   C_m} \sqrt {{a}^{2}-{y}^{2}}}{-b+a}}-2 {\frac {b{
   C_m} a\ln  \left( \sqrt {{a}^{2}-{y}^{2}}+a \right) }{-b+a}}
-2 { \frac {b{   C_m} \sqrt {{b}^{2}-{y}^{2}}}{-b+a}} \nonumber
\\
+2 {\frac {b{   C_m}
 a\ln  \left( \sqrt {{b}^{2}-{y}^{2}}+b \right) }{-b+a}}
 +2 {\frac {b {   C_m} c\ln  \left( \sqrt {{b}^{2}-{y}^{2}}+b
\right) }{-c+b}} \nonumber \\ -2 { \frac {b{   C_m} \sqrt
{{b}^{2}-{y}^{2}}}{-c+b}}  -2 {\frac {b{   C_m}
 c\ln  \left( \sqrt {{c}^{2}-{y}^{2}}+c \right) }{-c+b}}+2 {\frac {b
{   C_m} \sqrt {{c}^{2}-{y}^{2}}}{-c+b}}
\label{I_1l} \\
~ 0 \leq y < a \quad.  \nonumber
\end{eqnarray}

The second segment, $I^{II}(y)$, is
 \begin{eqnarray}
 I^{II}(y)=
-{\frac {b{   C_m} a\ln  \left( {y}^{2} \right) }{-b+a}}-2 {\frac
{b {   C_m} \sqrt {{b}^{2}-{y}^{2}}}{-b+a}}
\nonumber \\
+2 {\frac {b{   C_m} a\ln
 \left( \sqrt {{b}^{2}-{y}^{2}}+b \right) }{-b+a}}
 +2 {\frac {b{   C_m } c\ln  \left( \sqrt {{b}^{2}-{y}^{2}}+b
\right) }{-c+b}}
\nonumber \\
-2 {\frac { b{   C_m} \sqrt {{b}^{2}-{y}^{2}}}{-c+b}}-2 {\frac
{b{   C_m} c\ln
 \left( \sqrt {{c}^{2}-{y}^{2}}+c \right) }{-c+b}}
+2 {\frac {b{   C_m } \sqrt {{c}^{2}-{y}^{2}}}{-c+b}}
\label{I_2l} \\
  a \leq y < b  \quad. \nonumber
\end{eqnarray}
The third segment, $I^{III}(y)$, is
 \begin{eqnarray}
 I^{III}(y)=
 {\frac {b{   C_m} c\ln  \left( {y}^{2} \right) }{-c+b}}-2 {\frac
{b{    C_m} c\ln  \left( \sqrt {{c}^{2}-{y}^{2}}+c \right)
}{-c+b}} +2 { \frac {b{   C_m} \sqrt {{c}^{2}-{y}^{2}}}{-c+b}}
 \\
 b \leq y < c  \quad.  \nonumber
\label{I_3l}
\end{eqnarray}

\begin{figure*}
\begin{center}
\includegraphics[width=7cm]{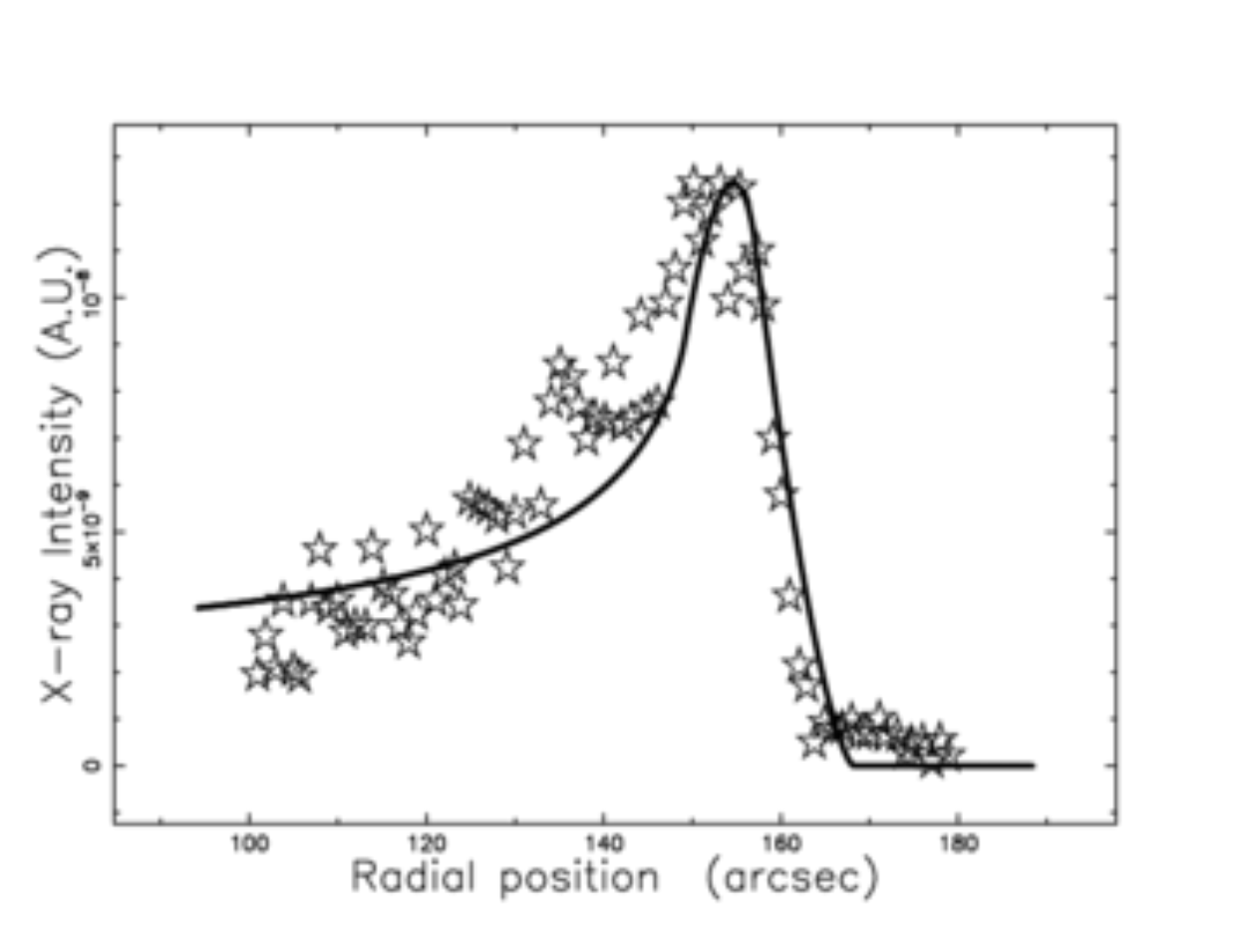}
\end {center}
\caption
{
 Cut of the mathematical  intensity ${   I}$
 of the diffusive  model, Eqs (\ref{I_1l}), (\ref{I_2l}) and  (\ref{I_3l}), 
 crossing the centre    (full  line) of  
 X-ray \s1006 
 observations. 
 The $x$- and  $y$-axes  are in arcsec,
 $a=149.14$, $b= 157$ arcsec,  $c=167.2 $ arcsec.
 The observed data  has been extracted  by the author
 from Fig.~8 (subfigure 14) of
 \cite{Ressler2014}.
}
\label{sn1006_diffusion}
    \end{figure*}

The profile of ${   I}$ which is  made by  three segments
(\ref{I_1l}), (\ref{I_2l}) and  (\ref{I_3l}), can be calibrated
against the real data of \s1006 as in 
  Fig.~8 (subfigure 14) of
 \cite{Ressler2014}   
and 
is reported in Figure~\ref{sn1006_diffusion}.

\section{Asymmetrical synchrotron radiation}

This section reviews a numerical algorithm which 
allows building a theoretical map for the intensity 
of radiation, and then applies 
the algorithm  to \sn1987a  and \s1006.

\subsection{The adopted algorithm}

We assume two  models  for the  luminosity:
a first non thermal model and a second thermal model. 
The source of synchrotron luminosity 
in the first model
is assumed here to be
the flux of kinetic energy, $L_m$,
\begin{equation}
L_m = \frac{1}{2}\rho A  V^3
\quad,
\label{fluxkineticenergy}
\end{equation}
where $A$ is the considered area, $V$ the velocity 
and $\rho$ the density, 
see formula (A28)
in \cite{deyoung}.
In our  case $A=R^2 \Delta \Omega$,
where $\Delta \Omega$ is the considered solid angle along 
the chosen direction.
This means
\begin{equation}
L_m = \frac{1}{2}\rho \Delta \Omega  R^2 V^3
\quad ,
\label{fluxkinetic}
\end{equation}
where $R$  is the instantaneous radius of the SNR and
$\rho$  is the density in the advancing layer
in which the synchrotron emission takes place.
The   observed luminosity along a given direction 
can  be expressed as
\begin{equation}
L  = \epsilon  L_{m}
\label{luminosity}
\quad  ,
\end{equation}
where  $\epsilon$  is  a constant  of conversion
from  the mechanical luminosity   to  the
observed luminosity in synchrotron emission.
The above  formula is derived in the framework
of the  `in situ' acceleration of electrons, see
formula (\ref{insitu}).
The numerical algorithm which allows us to
build  a complex  image will now be
outlined.
\begin{itemize}
\item  An empty (value=0)
memory grid  ${\mathcal {M}} (i,j,k)$ which  contains
$400^3$ pixels is considered.
\item  We  first  generate an
internal 3D surface by rotating the ideal image
 $180^{\circ}$
around the polar direction and a second  external  surface at a
fixed distance $\Delta R$ from the first surface. As an example,
we fixed $\Delta R = R/12 $, where $R$ is the
momentary  radius of expansion.
The points on
the memory grid which lie between the internal and external
surfaces are memorized on
${\mathcal {M}} (i,j,k)$ with a variable integer
number   according to formula
(\ref{fluxkinetic}) in the non-thermal case 
or to formula (\ref{eqn_transfer_square}) 
in  the thermal case.
\item  Each point of
${\mathcal {M}} (i,j,k)$  has spatial coordinates $x,y,z$ which  can be
represented by the following $1 \times 3$  matrix, $A$,
\begin{equation}
A=
 \left[ \begin {array}{c} x \\\noalign{\medskip}y\\\noalign{\medskip}{
   z}\end {array} \right]
\quad  .
\end{equation}
The orientation  of the object is characterized by
 the
Euler angles $(\Phi, \Theta, \Psi)$
and  therefore  by a total
 $3 \times 3$  rotation matrix,
$E$, see \cite{Goldstein2002}.
The matrix point  is
represented by the following $1 \times 3$  matrix, $B$,
\begin{equation}
B = E \cdot A
\quad .
\end{equation}
\item 
The intensity map is obtained by summing the points of the
rotated images
along a particular direction.
\item 
The effect of the  insertion of a threshold intensity, $I_{tr}$,
given by the observational techniques,
is now analysed.
The threshold intensity can be
parametrized  by  $I_{max}$,
the maximum  value  of intensity
characterizing the map.
\end{itemize}

\subsection{The case of \sn1987a} 

\begin{figure}
\includegraphics[width=6cm]{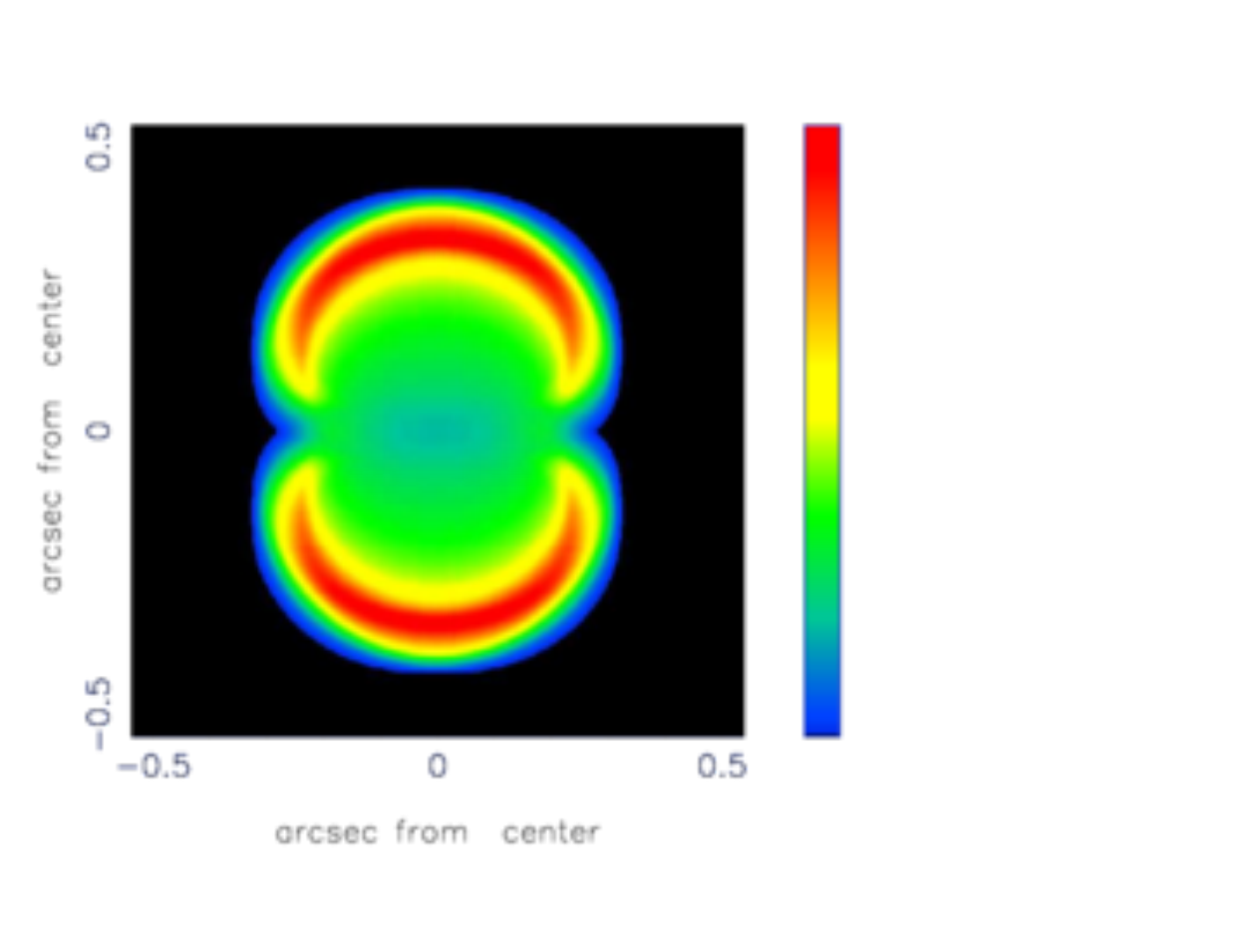}
\caption {
Map of the theoretical intensity  of
\sn1987a
in the presence of an hyperbolic density profile
with  parameters as in Figure~\ref{cut_hyper_1987a}.
The three Euler angles
characterizing the   orientation
  are $ \Phi $=180$^{\circ }$,
$ \Theta     $=90 $^{\circ }$
and   $ \Psi $=0  $^{\circ }$.
This  combination of Euler angles corresponds
to the rotated image with the polar axis along the
$z$-axis 
and thermal case}%
    \label{hyper_sn1987a_heat}
    \end{figure}

\begin{figure}
\includegraphics[width=6cm]{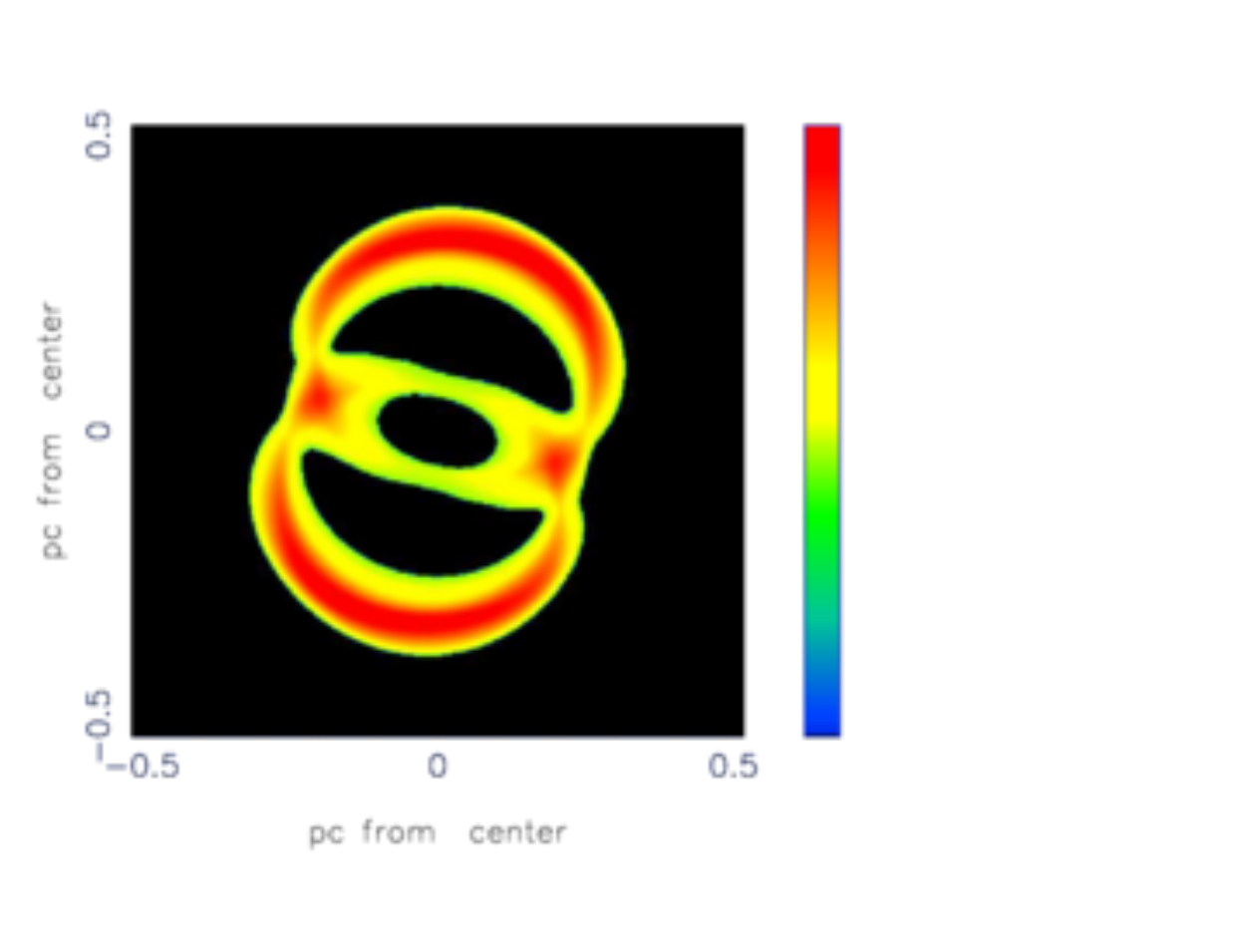}
\caption {
Model map of \sn1987a rotated in
accordance with the observations,
for an hyperbolic medium with
parameters as in Figure~\ref{cut_hyper_1987a}.
The three Euler angles
characterizing the   orientation
are
     $ \Phi   $=105$^{\circ }$,
     $ \Theta $=55 $^{\circ }$
and  $ \Psi   $=-165 $^{\circ }$.
This  combination of Euler angles corresponds
to the observed image and thermal case.
In this map $I_{tr}= I_{max}/2.2$ and
and thermal case%
}
    \label{hyper_sn1987a_hole}
    \end{figure}

The image of  \sn1987a
having  polar axis along the $z$-direction
is shown in Figure \ref{hyper_sn1987a_heat}.
The image  for a realistically rotated \sn1987a
is shown   in Figure \ref{hyper_sn1987a_hole}
and a comparison should be made
with  the observed triple ring system as reported
in Fig. 1 of \cite{Tziamtzis2011}
where the outer lobes are explained by light-echo
effects.
More details can be found in \cite{Zaninetti2013c}.

\subsection{The case of \s1006} 

\begin{figure}
\includegraphics[width=6cm]{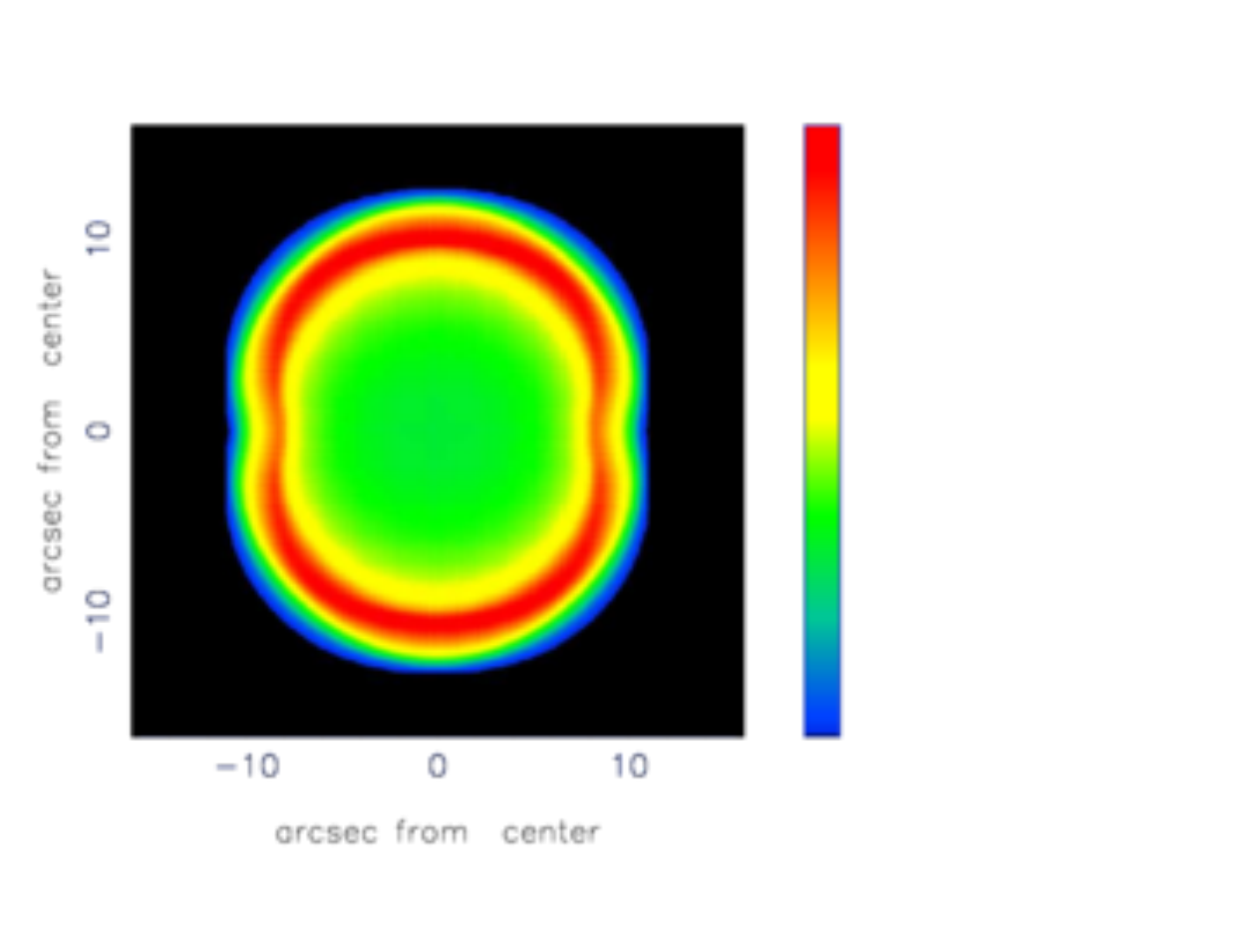}
\caption { 
Unrotated  map of \s1006   for an exponentially
varying medium and NCD case. 
Physical parameters as in Table \ref{datafit1006} 
and non-thermal case.
}
    \label{1006_heat}
    \end{figure}

\begin{figure*}
\includegraphics[width=6cm]{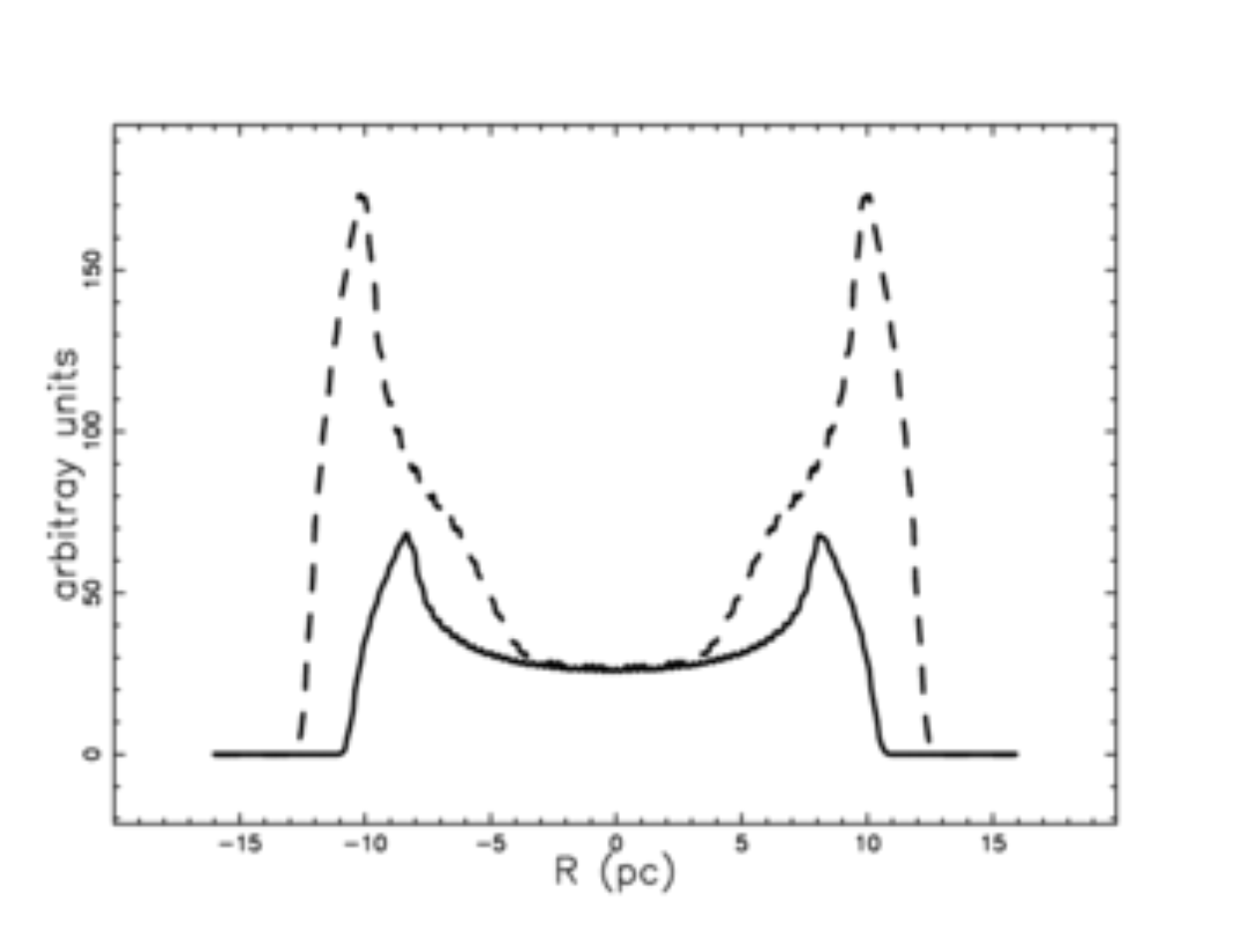}
\caption
{
 Two cuts along  perpendicular lines  of
 {   I }   for the unrotated image  of SNR  \s1006; 
non-thermal case
}
\label{cut_xy_1006_lum}
    \end{figure*}

\begin{figure*}
\includegraphics[width=6cm]{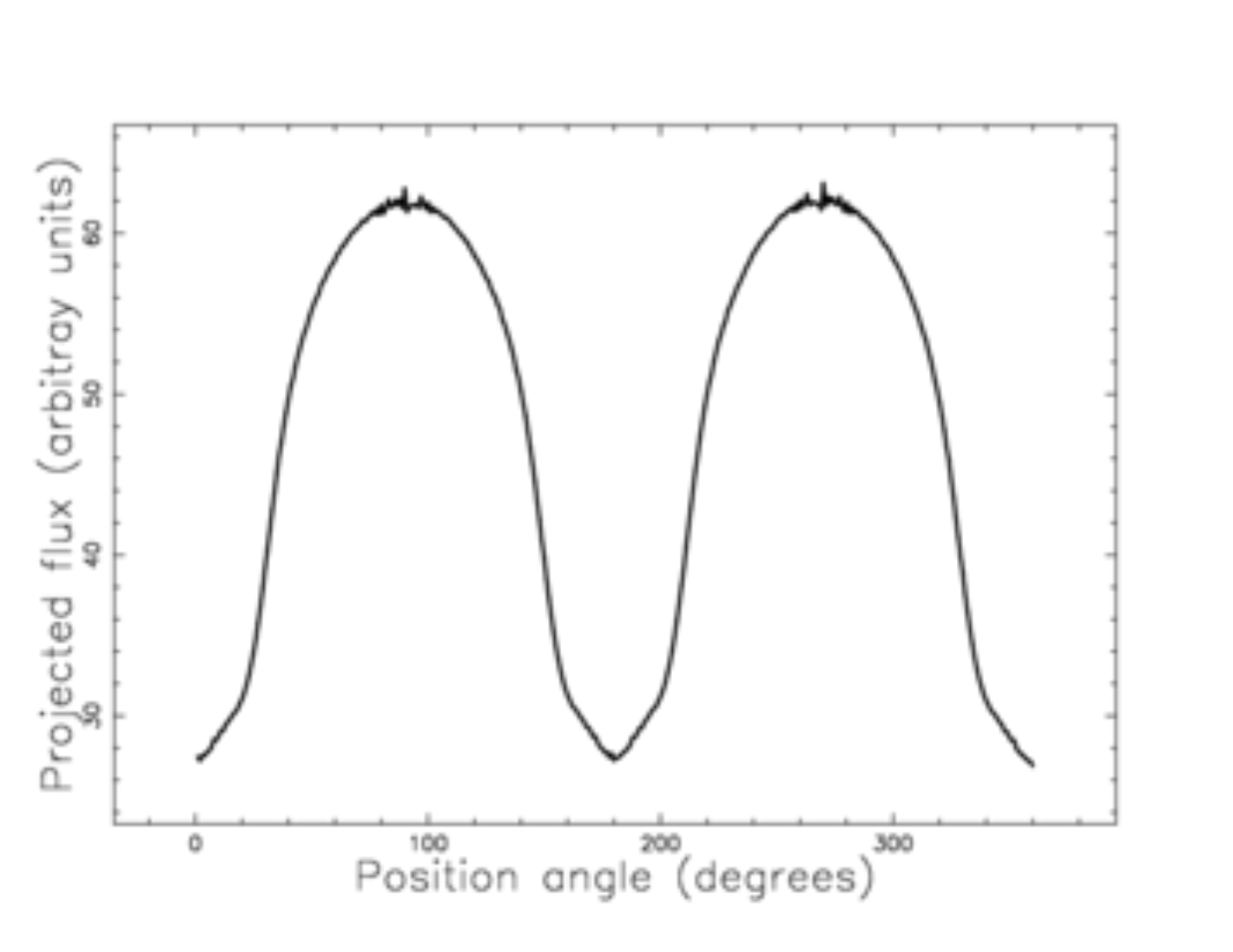}
\caption
{
Intensity
as function of the position angle
in degrees for SNR  \s1006; 
}
\label{flux_sn1006_360}
    \end{figure*}

The image of  \s1006   is  visible in different bands, such as
radio, see \cite{Reynolds1993,Reynoso2007}, optical, see
\cite{Long2007}, and X-ray, see  \cite{Dyer,Katsuda2010}. 
The  2D
map in intensity of \s1006 is  visible in Figure \ref{1006_heat}.
The  intensity along the equatorial and polar directions of our
image is reported in   Figure  \ref{cut_xy_1006_lum}; a comparison
should be made with Fig. 4 in \cite{Dyer}.
The  projected flux  as a function of the  position angle
is another interesting quantity to  plot,
see Figure  \ref{flux_sn1006_360},
and a comparison should  be made
 with  Fig. 5 top right
in  \cite{Rothenflug2004}.

\begin{figure}
\includegraphics[width=6cm]{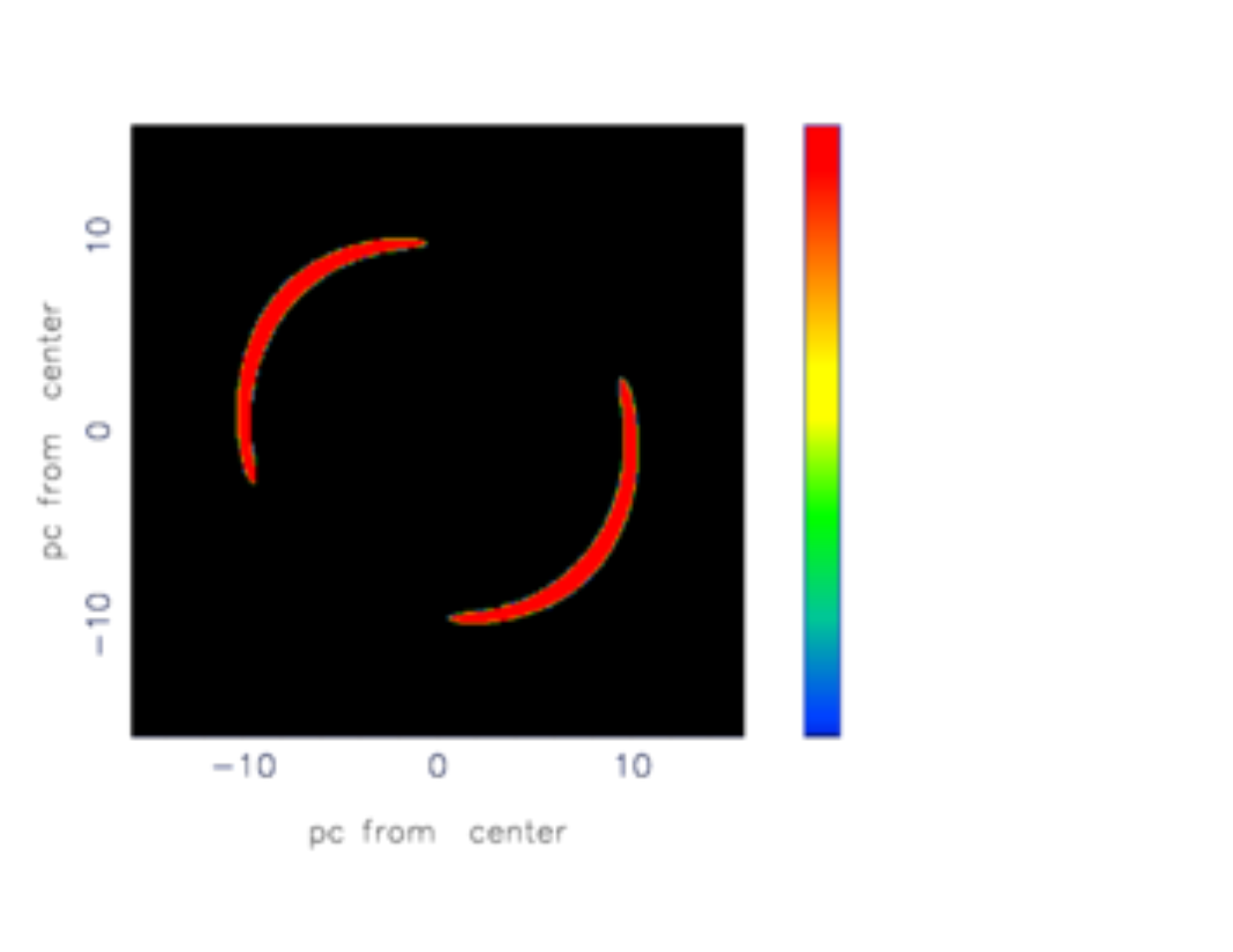}
\caption {
Model map of \s1006 rotated in
accordance with the X observations,
for an exponentially varying medium and NCD case.
Physical parameters as in Table \ref{datafit1006}.
The three Euler  angles characterizing
the orientation of the observer
are
     $ \Phi   $=90   $^{\circ }$,
     $ \Theta $=-55  $^{\circ }$
and  $ \Psi   $=-180 $^{\circ }$.
This  combination of Euler angles corresponds
to the observed image for 
the non-thermal case.
In this map $I_{tr}= I_{max}/1.1$.
          }%
    \label{1006_x}
    \end{figure}

After the previous graphs, it is  more simple to  present a
characteristic feature  such as  the "jet  appearance" visible in some
maps, see our  Figure \ref{1006_x}.
A  comparison of the above figure should be made
with 
the ASCA X-map in Figure \ref{sn1006asca}, see  nomenclature `Rim',  
the X-map  for the 6.33--6.53  keV  band  visible  in Fig. 3b of
\cite{Yamaguchi2008},
 the X-map  for the 2--7 keV band in Fig. 7 in \cite{Ressler2014},
and the X-map for 2--4.5 keV in Fig.~2   in \cite{Miceli2016}.

\section{Conclusions}

{\bf Type of medium:}

We have selected four density profiles, which decrease
with the distance ($z$-axis) from the equatorial plane.
The integral which evaluates the swept mass increases
in complexity according to the following
sequence of density profiles: hyperbolic,
power law,
exponential, and
Gaussian.

{\bf Classical thin layer}

The application of the thin layer approximation
with different profiles produces differential equations
of the first order.
The solution of the first-order differential equation
can be  analytical in the classical case, characterized by a
hyperbolic density profile,
see (\ref{rtanalyticalhyper}),
and numerical  in all other cases.
We also evaluated the approximation of the solution
as a power law series,
see
(\ref{rtseriesexp}),
or using the  Pad\'e approximant, see (\ref{rtpade}):
the differences between the two approximations
are outlined in Figure \ref{pade_exp}.
In the case of  an exponential density profile
for the CSM
as given by equation (\ref{profexponential}),
the NCD approximation was implemented.
The presence  of a gradient  for the density  transforms
a spherical symmetry  into an axial  symmetry
and allows  the appearance  of
the so called `bipolar motion'.

{\bf Relativistic thin layer}

The application of the thin layer approximation
to the relativistic case produces first-order differential
equations which can be solved only numerically
or as a power series,
see
(\ref{rtserieshyperrel}).
and
(\ref{rtseriesexprrel}).

{\bf Symmetric image}

The intensity of the image of a symmetrical SN  or SNR 
in the case of an optically thin medium can be computed
through
an analytical evaluation of the lines of sight 
when the number density is constant between two spheres,
see formula~(\ref{irim}).
In the case  of  a  symmetrical 
diffusive  process  which  is  built in 
presence  of three  spheres, 
the 
intensity of emission  is assumed to be proportional
to  the number density, see  
formulas (~\ref{I_1l}), (~\ref{I_2l}) and  (~\ref{I_3l}).

{\bf Asymmetric image}

A first model for the emissivity in the advancing  layer assumes
a   proportionality   to the  flux of  kinetic energy,
see  equation   (\ref{fluxkineticenergy})
where  the density  is assumed to be
proportional  to the swept material.
A second  model assumes an emissivity 
proportional to the square of the number density,
see  formula (\ref{eqn_transfer_square}).
This  second  allows simulating
particular effects, such as
the triple ring system of \sn1987a, see
Figure  \ref{hyper_sn1987a_hole}.
Another  curious  effect is  the  "jet  appearance" visible in the
weakly  symmetric \s1006, see  Fig. \ref{1006_x};
this jet's effect is simulated in the framework of 
the first non thermal model. 
The jet/counter
jet effect plays a significant role in current research, see
the discussion in Section 5.2 of \cite{Dopita2006}, where the  jet
appearance is tentatively explained by neutrino heating, see
\cite{Walder2005}, or  by the MHD jet, see \cite{Takiwaki2004}.
Here in contrast we explain the  appearance  of the jet by the
addition of three effects:
 \begin{itemize}
\item  An asymmetric law of expansion due to a density gradient 
with respect to the equatorial plane. This   produces an asymmetry in
velocity. 
\item  The direct conversion of the flux of kinetic
energy into radiation.
 \item  The image  of the SNR as the composition of integrals
 along the line of sight.
 \end{itemize}
  A careful  calibration  of the various  parameters
  involved  can be done  when  cuts  in intensity  are available.

\section*{Acknowledgments}

Credit for Figure  
\ref{sn1987a_st} 
 is  given to  
 the  Hubble Space Telescope,
for Figure  
\ref{SN_CHANDRA_X_2013} is given to 
the Chandra X-ray Observatory,
and for Figure  
\ref{sn1006asca} is given to 
the ASCA X-ray Observatory.
In this research we have used
WebPlotDigitizer made by Ankit Rohatgi
and  available at \url{https://automeris.io/WebPlotDigitizer}.


\end{document}